\documentclass[apj,twocolumn]{emulateapj}

\newfont{\Sc}{eusm10}

\shorttitle{Lyman $\alpha$ Protocluster at $z=3.1$}
\shortauthors{Yamada et al.}

\begin{document}

\title{Panoramic Survey of Lyman $\alpha$ Emitters at $z=3.1$}

\author{T. Yamada, \altaffilmark{1}, Y. Nakamura\altaffilmark{1}, Y. Matsuda\altaffilmark{2}, T. Hayashino\altaffilmark{3}, R. Yamauchi\altaffilmark{3},N. Morimoto\altaffilmark{1}, K. Kousai\altaffilmark{3}, M. Umemura\altaffilmark{4} }
\email{yamada@astr.tohoku.ac.jp}

\altaffiltext{1}
{Astronomical Institute, Tohoku University, Aramaki,
Aoba-ku, Sendai, Miyagi, 980-8578}
\altaffiltext{2}
{Department of Physics, Durham University, South Road, Durham DH1 3LE}
\altaffiltext{3}
{Research Center for Neutrino Science, Graduate School of Science, Tohoku Univsersity, Sendai 980-8578, Japan}
\altaffiltext{4}
{Center for Computational Physics, University of Tsukuba, Tsukuba, Ibaraki 305, Japan}

\begin{abstract}

 We present the results of the extensive narrow-band survey of Ly$\alpha$ emission-line objects at $z$=3.1 in the 1.38 deg$^2$ area surrounding the high density region of star-forming galaxies at $z$=3.09 in the SSA22 field, as well as in the 1.04 deg$^2$ area of the three separated general blank fields. In total of 2161 Ly$\alpha$ emitters, 1394 in the SSA22 fields and 767 in the general fields, respectively, are detected to the narrow-band $AB$ magnitude limit of 25.73, which corresponds to the line flux of $\approx$1.8$\times 10^{-17}$ erg s$^{-1}$ cm$^{-2}$ or luminosity of $\approx$1.5$\times 10^{42}$ erg s$^{-1}$ at $z=3.1$, above the observed equivalent width threshold, $\approx$190\AA . The average surface number density of the emitters at $z$=3.1 in the whole general fields above the thresholds is 0.20$\pm 0.01$ arcmin$^{-2}$. The SSA22 high-density region at $z$=3.09 whose peak local density is 6 times the average is found to be the most prominent outstanding structure in the whole surveyed area and is firmly identified as a robust `protocluster' with the enough large sample. We also compared the overdensity of the 100 arcmin$^2$ and 700 arcmin$^2$ areas which contain the protocluster with the expected fluctuation of the dark matter as well as those of the model galaxies in cosmological simulations. We found that the peak height values of the overdensity correspond to be 8-10 times and 3-4 times of the expected standard deviations of the counts of Ly$\alpha$ emitters at $z$=3.1 in the corresponding volume, respectively. We conclude that the structure at $z$=3.09 in the SSA22 field is a very significant and rare density peak up to the scale of $\approx 60$ Mpc.

\end{abstract}

\qquad

\section{Introduction}

  Ly$\alpha$ emission from high-redshift galaxies is an important and useful probe in studying the star formation and structure formation history of the universe (Partridge \& Peebles 1967). The sensitive narrow-band imaging observations have succeeded to detect a large number of the Ly$\alpha$ emitters at intermediate and high redshift toward $z \sim 7$ (e.g., Hu \& McMahon 1996; Steidel et al. 2000; Rhoads and Malhotra (2001); Rhoads et al. (2003); Kodaira et al. 2003; Palnus et al. 2004; Hayashino et al. 2004; Matsuda et al. 2004; 2010; 2011; Taniguchi et al. 2005; Gronwall et al. 2007; Iye et al. 2006, Ouchi et al. 2008; 2010). 

 As the narrow-band imaging can constrain the redshift of the objects to the small range, Ly$\alpha$ emitters are also useful in studying the spatial distribution of the star-forming galaxies at high redshift. It is known that high-redshift galaxies are strongly clustered due to the formation bias to the underlying mass distribution (Kauffman et al. 1999; Benson et al. 2001; Weinberg et al. 2004). While it is not trivial that Ly$\alpha$ emitters follow the simple halo-mass bias in the hierarchical structure formation, since the origins and time-scale of the Ly$\alpha$ emission may be diverse, the high-density region of the Ly$\alpha$ emitters may mark the region where galaxy formation preferentially occurs at the epoch. Typical spectral resolution $\delta\lambda / \lambda$ of the narrow-band imaging is 50-100, which corresponds to the comoving scale of 50-100 Mpc at $z \sim 3$, which is narrow enough to locate the clustered regions and voids.

 More than a few high-density regions of the Ly$\alpha$ emitters and star-forming galaxies at high redshift have been studied (Keel et al. 1999; Steidel et al. 2000; Shimasaku et al. 2003; Palnus et al. 2004; Hayashino et al. 2004; Matsuda et al. 2004; 2009; 2010; Ouchi et al. 2005; Steidel et al. 2005; Kajisawa et al. 2006; Miley et al. 2006; Venemans et al. 2007; Hatch et al. 2008; 2009; Overzier et al. 2008; Digby-North et al. 2010; Kuiper et al. 2010; Yang et al. 2010). Among them, the structure at $z$=3.09 in the SSA22 region, namely the area around the original selected survey area at 22h (Cowie et al. 1990) is one of the most prominent and the interesting ones. The large density excess was first discovered in the redshift distribution of the Lyman Break galaxies (LBGs) by Steidel et al. (1998). The density of the region within 20$\times$15$\times$21 Mpc$^3$ (in the flat $\Lambda$CDM with $H_0$=70 km s$^{-1}$Mpc$^{-1}$ ) comoving volume was reported to be 6 times of the average. In fact, the peak has the largest density enhancement in their whole survey volume  (Adelberger et al. 1998; Steidel et al. 2003) and it is still one of the largest density peaks discovered at high redshift so far. Subsequent narrow-band observations (Steidel et al. 2000; Hayashino et al. 2004; Matsuda et al. 2004) found that the Ly$\alpha$ emitters, which have some overlap with LBGs but clearly the different sample in terms of the UV luminosity and the Ly$\alpha$ equivalent width, also show the similar density enhancement. The following multiwavelength studies of the field show that the density of the Active Galactic Nuclei (AGN) detected in X-ray as well as the infrared or sub-mm sources also high compared to other fields (Geach et al. 2005; 2007; Lehmer et al. 2009; Webb et al. 2009; Tamura et al. 2009). 

\figurenum{1}
\begin{figure}[!t]
\includegraphics[width=8cm]{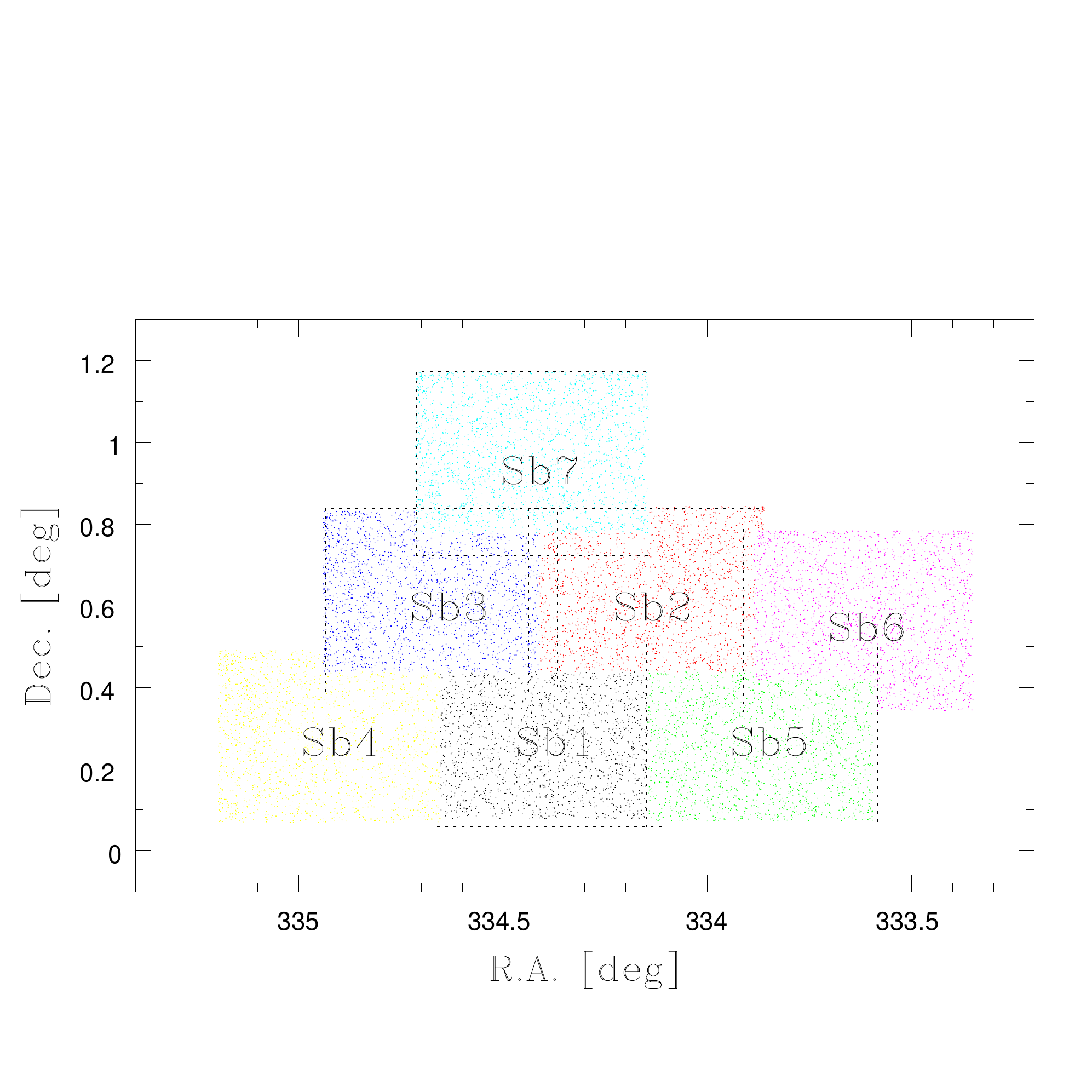}
\caption{The observed Suprime Cam field of views of the SSA22 fields. The boundary used for the analysis in this paper is shown by the colors of the dots which are the $NB497$-detected objects sampled with the rate of one in 50.}
\label{fig1}
\end{figure}

 While the region is often referred as a `protocluster', there has been still uncertainty in the value of the density enhancement of the Ly$\alpha$ emitters as a distinguished structure. Firstly, either LBGs or Ly$\alpha$ emitters does not show the prominent concentration on the sky and their distribution rather extends over the observed field of view of Steidel et al. (1998) or of Steidel et al. (2000) although some local density structure is recognized (Steidel et al. 2000). Hayashino et al. (2004) indeed revealed that the region is a part of the `belt-like' large-scale structure with the comoving scale of $\sim 30-50$ Mpc which also extends over the observed field of Subaru Suprime Cam, $\approx 27'$$\times$ 34$'$. While the subsequent spectroscopic observation revealed that this large-scale high-density structure seems to consist of the filamentary structures in the redshift space and the density peak discovered by Steidel et al. may locate near the intersection of the filaments (Matsuda et al. 2005), further observations covering the larger area is definitely needed to characterize the density peak in the enough large volume.  
 The control sample should also be improved. When Steidel et al. (2000) evaluated the density enhancement of Ly$\alpha$ emitters, the compared `field' sample were those obtained earlier in a relatively small volume (Cowie and Hu 1998; Pascarelle et al. 1996; 1998). In Hayashino et al. (2004), they referred the sample obtained by themselves in a general field of the similar area as the SSA22-Sb1. The data was, however, shallower than that obtained at the SSA22 field and the significance of the overdensity was evaluated by using only the relatively bright objects. Single field for the control sample is also not enough to evaluate the field-to field variation of the number density of the emitters.

 In this paper, we present the results of our wide-field and deep surveys of the Ly$\alpha$ emitters at $z$=3.1 around the dense structure in the SSA22 fields (Steidel et al. 1998; 2000; Adelberger et al. 1998; Hayashino et al. 2004) as well as the general blank deep survey fields. The survey volume is significantly enlarged to be 10 times larger than that in Hayashino et al. (2004). The  goal of this paper is to characterize the high-density region of the Ly$\alpha$ emitters and to evaluate the significance of their density enhancement. The detection of the extended Ly$\alpha$ nebulae, namely, Ly$\alpha$ blobs in the dame fields  was reported in the previous paper, Matsuda et al. (2011)

%
%

\figurenum{2}
\begin{figure}[!h]
\includegraphics[width=75mm]{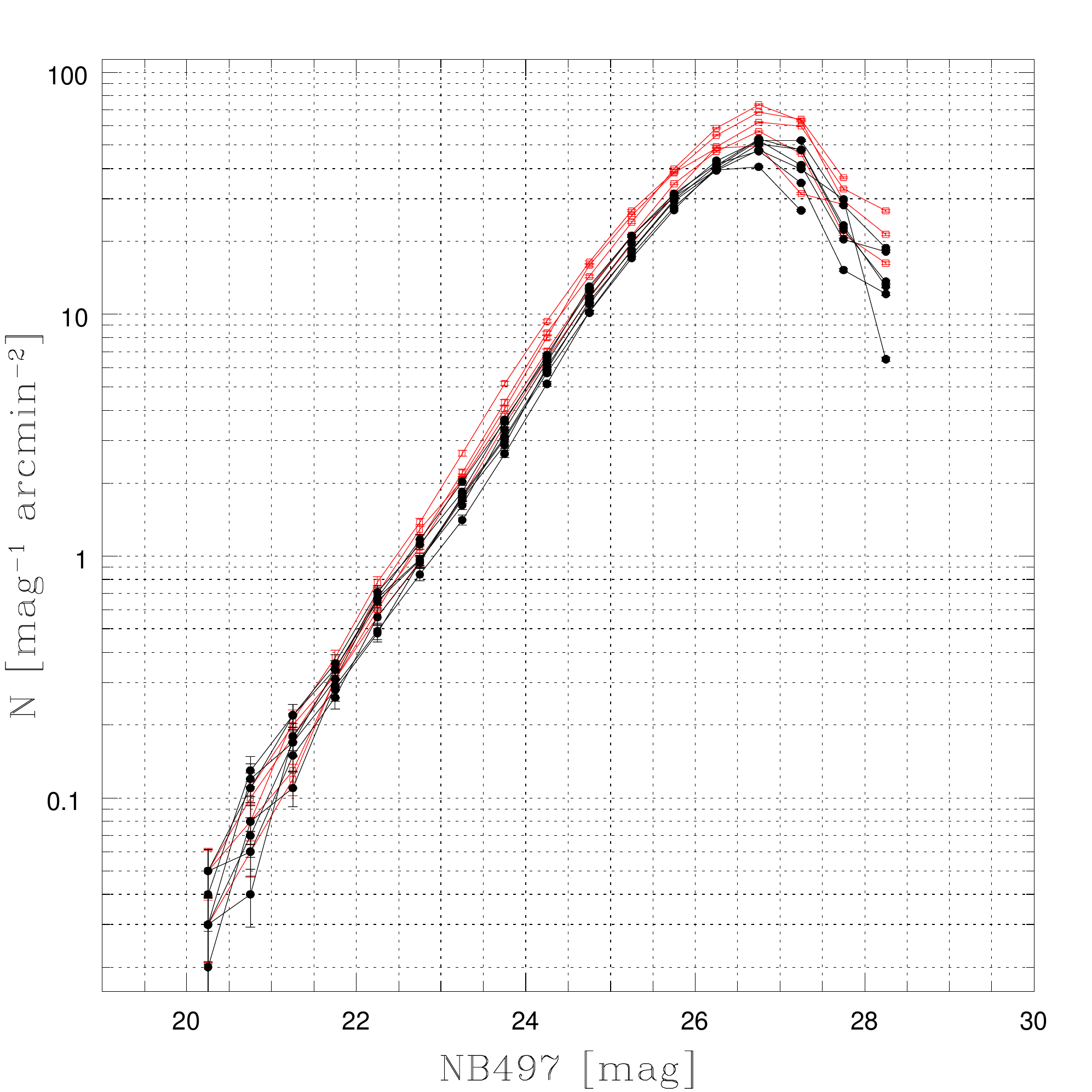}
\caption{\small The observed number counts of the sourced setected in the $NB497$ images. The data is not corrected for Galactic extinction. The filled circles and black lines show the data for the SSA22 Sb1-Sb7 fields and the open squares and red lines show those of the general fields.}
\label{fig2}
\end{figure}

\section{Data}

\subsection{Observations}

  We conducted the deep and wide-field imaging observations with Subaru 8.2m telescope equipped with the Subaru Prime Focus Camera (Suprime Cam) (Miyazaki et al. 2002). The custom narrow-band filter $NB497$ (Hayashino et al. 2004) whose central wavelength and width are 4977\AA\ and 77\AA , respectively, was used to sample the Ly$\alpha$ emitters at $z$=3.09. The width corresponds to the redshift range $z$=3.062$-$3.125, or 59 Mpc in comoving scale. 

 In order to expand the $z$=3.1 Ly$\alpha$ emitter survey around the SSA22 region, we newly observed the six adjacent Suprime Cam field of views (FoV) around the original field, SSA22-Sb1 (Hayashino et al. 2004; Mtasuda et al. 2004). The coordinates of the fields, referred as SSA22-Sb2 to Sb7, and the summary of the observations are shown in Table 1.  Their configuration on the sky is shown in Fig.1. The boundaries of the fields used in this paper are indicated by the colors of the dots which show the positions of the detected sources (one in the fifty objects are plotted). Typical exposure time for the $NB497$, $B$, and $V$-band filters are 5 hours, 1 hour, and 1 hour, respectively. The seeing condition was fairly homogeneous except for those in some observations of SSA22-Sb4 field where we need to apply a slight aperture correction to match with the other data (see below).

 For the control sample, we also obtained the similar deep narrow-band data in the five general blank fields, namely Subaru/XMM-Newton Deep Survey (SXDS)-North, -Center, -South (Furusawa et al. 2008), Subaru Deep Field (SDF) (Kashikawa et al. 2004), and the field around Great Observatory Optical Deep Survey North (GOODS-N) (Dickinson et al. 2004) fields. The broad-band data of the SXDS and SDF fields taken and published by each deep survey team was used. As the available public broad-band data of Suprime Cam in the GOODS-N fields is not as homogeneous as the other general fields, we have to be careful in analyzing the equivalent width distribution of the GOODS-N field. We here simply omitted it in the discussion of the equivalent width. 

 Thus, in this paper, the data for in total of 7 Suprime Cam fields in the SSA22 region including the original SSA22-Sb1 field as well as those of the 5 Suprime Cam fields in the three separated general blank fields (SXDS-N, -C, and -S, SDF, GOODS-N) are analyzed. The total areas used in the analysis are 1.38 deg$^2$ for the SSA22 fields and 1.04 deg$^2$ for the blank fields, respectively.

\subsection{Data Reduction}

 The data reduction was done by using SDFRED (Yagi et al. 2002; Ouchi et al. 2004). The basic procedure is the same as described in Hayashino et al. (2004). After subtracting the {\it bias} by using the mean value of the overscaned data, the frames are flat-fielded by the flat-fielding frames constructed by the night sky images. The instrumental image distortion and the differential atmospheric dispersion were also corrected. Sizes of the point spread function (PSF) are then measured by the bright stars and the images are smoothed to be matched with the target PSF size. We match all the images except for SSA22-Sb4 and GOODS-N to the full width of half maximum (FWHM) of 1.$''$0. Since some of the images of SSA22-Sb4 and GOODS-N were relatively poor in seeing, we smoothed them to the FWHM of 1.$''$1. The sky background was then subtracted by the standard procedure in SDFRED. We adopted the box of 72 pixels $\times$ 72 pixels for the local sky subtraction. After masking the bad regions, all the chip images are properly aligned and median combined. The broad band images were also aliened to the narrow-band images. Astrometry was done by using the USNVO stars.

 The absolute photometric calibration was done by using the spectroscopic standard stars for the $NB497$ images at SSA22-Sb1 (Hayashino et al. 2004) and at the general fields, and the Landolt standard stars for the $B$ and $V$ band images at SSA22-Sb1. For the other SSA22 fields Sb2-Sb7, we relatively calibrate them to the Sb1 image by using the overlapping area to ensure the relative homogeneity over the whole area.  The uncertainty of the absolute photometric calibration is as large as 0.1 mag but the relative one among the SSA22 fields is 0.01 mag or less at the magnitude range $m_{AB}$=25-26 which cover the major fraction of the Ly$\alpha$ emitters. Taking that the broad-band magnitude values listed in the published catalog of SDF and SXDS objects are robust, we also added a small correction ($\sim 0.05-0.1$ mag) to the absolute zero point values (after the correction for Galactic extinction as described below) so that the relative calibration between the SSA22 fields and general fields is also accurate enough to make the homogeneous sample. We use the $AB$ magnitude system through the paper and the magnitude in the narrow-band is also referred as $`$$NB497$'.

%
%

\figurenum{3}
\begin{figure}[!t]
\includegraphics[width=8cm]{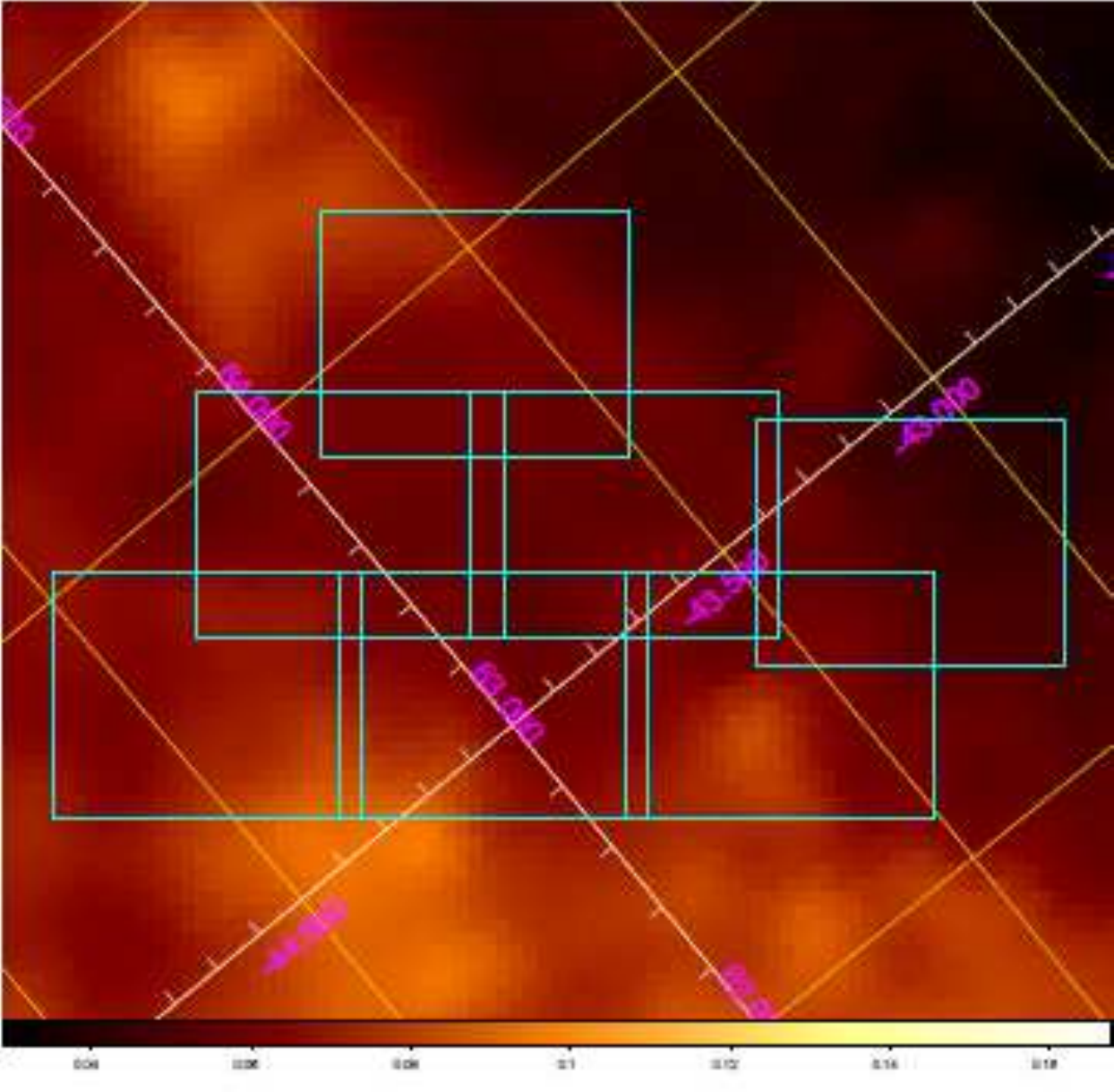}
\caption{The extinction map converted from the IRAS 100 $\mu$m data in Schlegel et al. (1998) of the SSA22 fields. The color bar at the bottom show the $E$($B-V$) value scale. The map is taken from NASA/IPAC Infrared Service Archive http://irsa.ipac.caltech.edu/applications/DUST/. The location of the SSA22-Sb1 to Sb7 are shown by the cyan boxes and the Galactic coordinate grid is also overplotted}
\label{fig3}
\end{figure}

\subsection{Photometry and the Galactic extinction correction}

 The sources are detected by using the software SExtractor 2.1.6 (Bertin and Aunouts 1996). The combined image of each field in each filter was analyzed with SExtractor and the pseudo total flux, FLUX AUTO (or MAG AUTO for magnitude) within 2.5 $\times$ Kron radius as well as the flux within $2''$-diameter circular aperture were obtained for the detected objects. We also analyzed the broad-band images together with the narrow-band images by using the SExtractor double-image mode and the FLUX AUTO within the Kron aperture as well as the fixed aperture at the position of the narrow-band detected sources were obtained. The photometric errors are dominated by the sky background noise and the 5$\sigma$ limiting magnitude for the $2''$-diameter aperture in each field are shown in Table 1. Sometimes the narrow-band-selected objects are too faint to be detected significantly in the continuum image. In calculating and analyzing the colors or the equivalent width of the objects we will substitute them as the 1$\sigma$ limiting magnitude for the aperture. 

 To obtain the appropriate continuum flux to be compared with the narrow-band flux, we constructed the weighted `$BV$-band' images by summing the $B$ and $V$-band images by $(2B+V)/3$. We refer this average magnitude as $BV$.

 The seeing of the SSA22-Sb4 images was relatively poor and we could not match them to the target image size of $1.''0$. As the seeing in other ten fields is better, in spite that we match all the images to the poorest SSA22-Sb4 one, we made a small correction, 0.061, 0.065, 0.070, and 0.069 mag to the aperture magnitude in $B$, $NB497$, $V$, and $BV$ band for the sources in SSA22-Sb4, respectively. These correction values were obtained by comparing the observed and the artificially degraded images of SSA22-Sb1 and confirmed by comparing the sources in the Sb1-Sb4 overlap region. Since the broad-band data of the GOODS-N field has also relatively poor image quality, we also smoothed all the GOODS-N-field images to FWHM=1.$''$1 and applied the same aperture correction as for SSA22-Sb4.
 
 Fig.2 shows the number counts in $NB497$ band of the detected sources in the SSA22 fields (filled circles and the black lines) as well as for those in the general fields (open squares with the red lines). The counts in the SSA22 fields are systematically {\it low}, which is due to the Galactic extinction as the SSA22 fields locates at relatively low Galactic latitude, ($l$, $b$) $\approx$ (63$^\circ$, $-$45$^\circ$). We therefore have to correct the flux of the objects in the SSA22 fields to the Galactic extinction in order to compare the data with the general fields which are at the higher Galactic latitude. For example, the mean extinction in $V$ band is 0.22 mag in SSA22 while they are 0.08 and  0.06 mag in the SXDS and SDF fields, respectively. The difference of the extinction within the seven SSA22 Suprime Cam fields should also be carefully considered as we will discuss the inhomogeneity in the spatial distribution of the Ly$\alpha$ emitters over the fields. For the purpose, we refer the most modern map of Galactic reddening over the sky based on the {\it COBE} DIRBE and {\it IRAS ISSA} far-infrared data (Schlegel, Finkbeiner, \& Davis 1998, hereafter SFD) shown in Fig.3 with the projected seven Suprime Cam FoVs. The spatial resolution of the SFD map is 0.32$^\circ$. The color excess $E$($B-V$) ranges from 0.05 to 0.09 gradually changes from northwest to southeast. 

%
%
\figurenum{4}
\begin{figure}[!h]
\includegraphics[width=78mm]{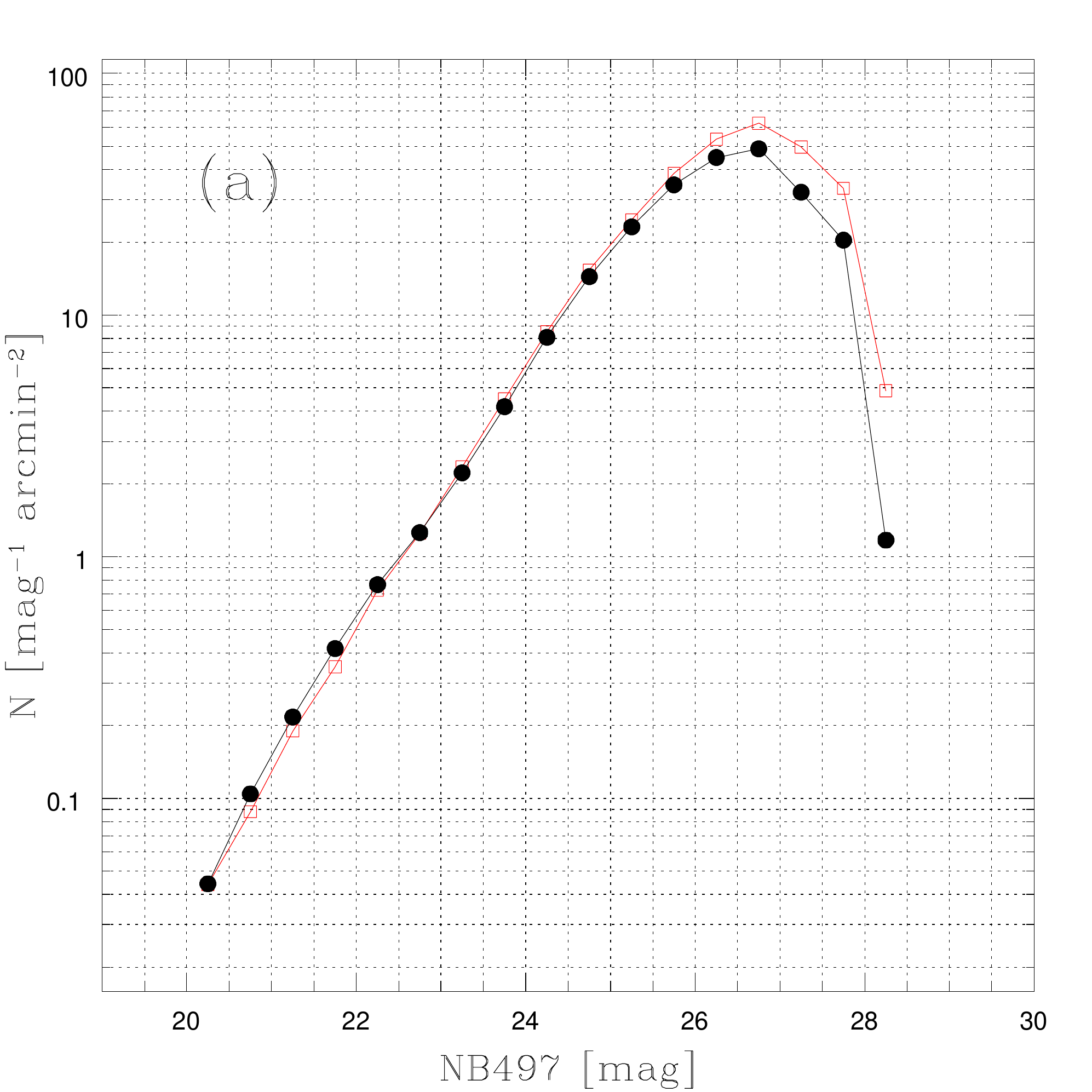}
\caption{(a) The extinction corrected numbour counts of the averages of the SSA22 (black line and simbols) and the general fields (red one).}
\label{fig4a}
\end{figure}
\figurenum{4}
\begin{figure}[!h]
\includegraphics[width=78mm]{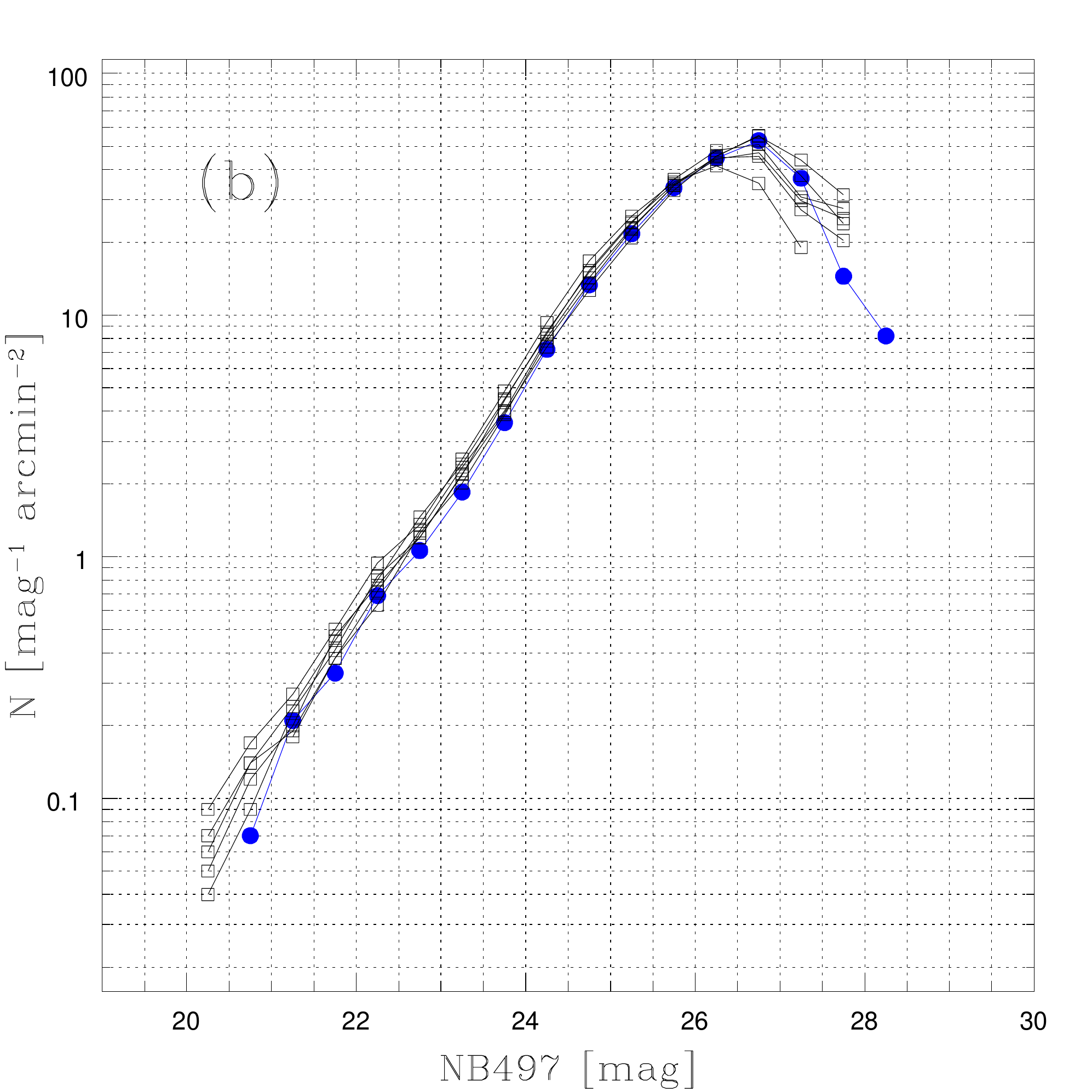}
\caption{(b) Same as (a) but those for the each seven SSA22 fields. The counts of SSA22-Sb1 field is shown by the blue line. The shallowest data is that of SSA22-Sb4 field.}
\label{fig4b}
\end{figure}

 For all the detected sources we applied the correction for Galactic extinction from the $E(B-V)$ value at the position of the SFD map by using the following relations, $A_V$ = $3.16 \times$$E$($B-V$), $A_B$=$4.11 \times$$E$($B-V$), and $A_{NB497}$=$3.55 \times$$E$($B-V$), based on the extinction curve derived by Kenyon et al. (1994). Fig.4a and 4b  show the comparison of the source counts between the SSA22 and the general fields as well as those among the seven SSA22 fields {\it after} the Galactic extinction correction. While the SFD map smoothed out the patchy structure smaller then 0.32$^\circ$, the agreement between the average of the corrected counts in the SSA22 fields and the general fields is reasonably good. The scatter among the seven SSA22 fields also becomes small after the correction while the field-to-field variation with $\sim 10$$\%$ in the number density remains. Note that the counts in SSA22-Sb1 field, shown by the blue line, is the lowest among them and it would not to be the any artificial cause of the overdensity of the Ly$\alpha$ emitters in the field. 

 Fig.5 shows the distribution of the $BV-NB497$ colors for the objects between 23 mag and 25 mag in both the bands. The vertical dashed (red) line indicates the peak (mode) value of the SSA22-Sb1 field and the dotted line in each panel shows that of the each field. The peak coincides within $\sim 0.02$ mag, which enable us to make the uniform selection of the narrow-band excess objects.

%
%

\figurenum{5}
\begin{figure}[!t]
\includegraphics[width=8cm]{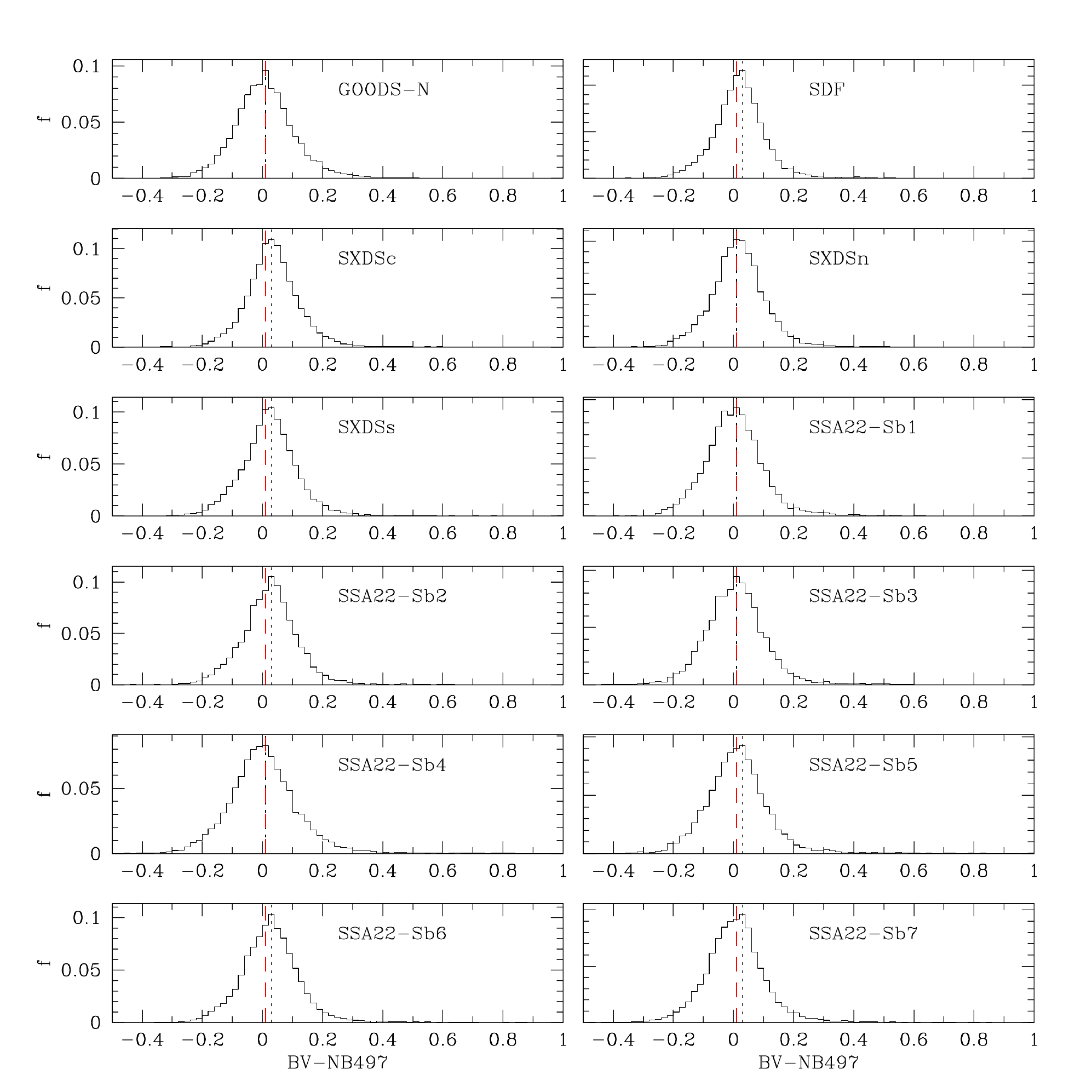}
\caption{The distribution of the $BV-NB497$ color for the objects between 23 mag and 25 mag in both the bands.}
\label{fig5}
\end{figure}

\section{Identification of the Ly$\alpha$ Emitters}

 We now describe the selection criteria for the $z$=3.1 Ly$\alpha$ emitters. The aperture magnitude values are used for the purpose. Fig.6 and Fig.7 show the color-magnitude diagram of the objects detected in the $NB497$ images in the SSA22 fields and in the general fields, respectively. The further very slight correction for the color zero point left after the photometric calibration above was made in the Ly$\alpha$ emitter identification so as to ensure more complete sample homogeneity over the fields.

 The criteria to identify the Ly$\alpha$ emitters at $z \sim 3.1$ are as follows.

 ${\rm (1)} \quad NB497 < 25.73  \quad {\rm (S/N > 6.6)}$, and, 

 ${\rm (2a)} \quad BV-NB497 > 1.0  \quad \& \quad  B-V_c > 0.2$ 
 
\qquad \qquad (when  $V_c < 26.5$),

 or, 

 ${\rm (2b)} \quad BV-NB497 > 1.3 $

\qquad \qquad (when $V_c > 26.5$).

$V_c$ is the line-corrected $V$-band magnitude. The criteria are similar as in Hayashino et al. (2004) but the color threshold was slightly relaxed based on the fact that few contaminations by the foreground objects were found in our previous spectroscopic observations.  In fact spectroscopic observations (Matsuda et al. 2005; 2006) shows that the contamination of the foreground objects, mostly [OII] emitters at $z$=0.33, is at most 1$\%$, or negligible. The $B-V_c$ criteria is useful to further prevent the contamination of the $z$=0.33 [OII] emitter although we cannot apply it to the continuum faint sources, $V_c > 26.5$. 

 For the robustness of the sample, we also apply an additional constraint that the observed $BV-NB497$ color must be larger than the 4$\sigma$ value of the background noise in the SSA22-Sb4 field. The limit is shown by the curves in Fig.6 and 7. Only a small fraction of the objects are rejected by the last criteria.

%
%
\figurenum{6}
\begin{figure*}[!t]
\includegraphics[width=16cm]{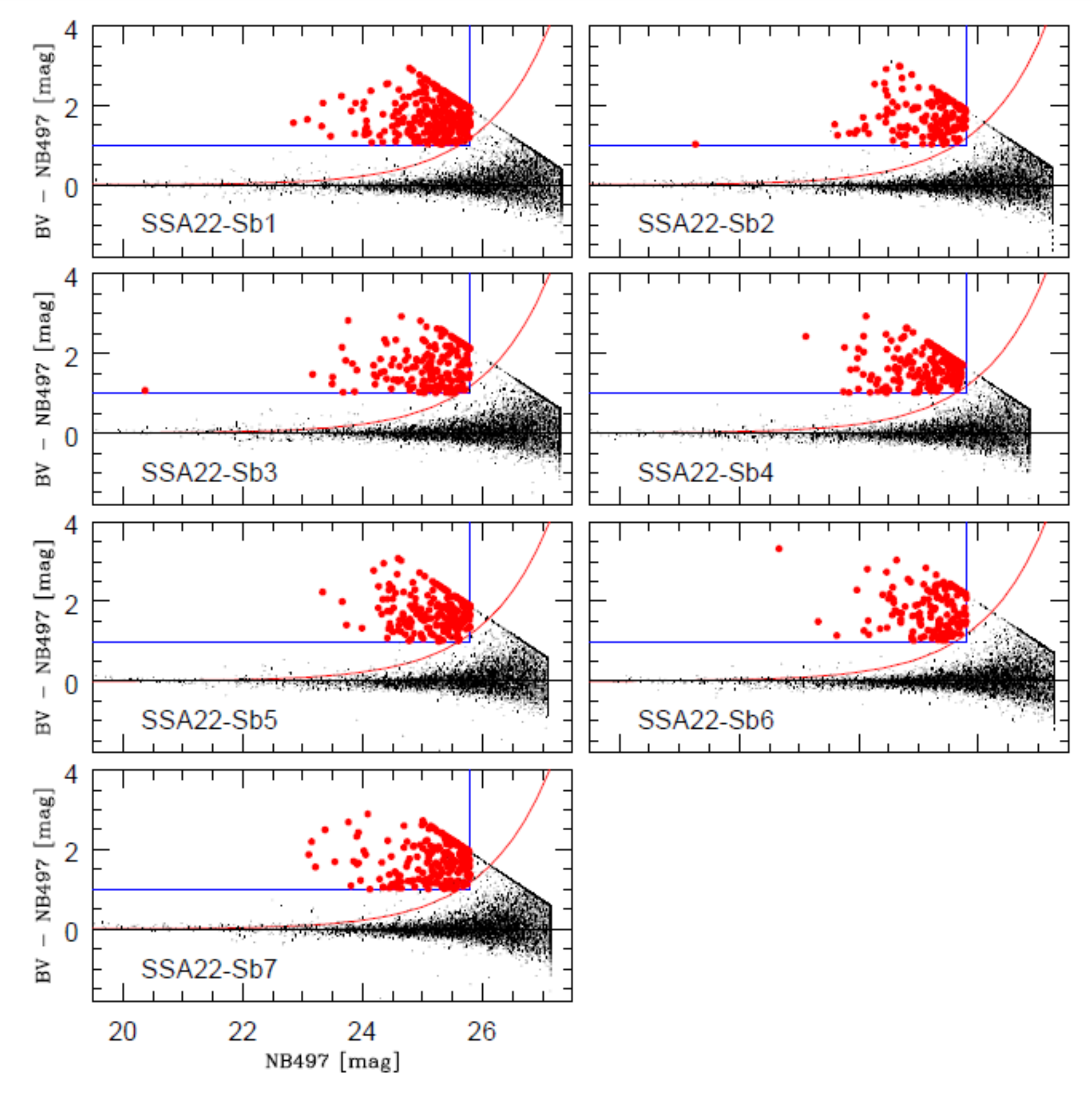}
\caption{The $BV$-$NB497$ and $NB497$ color magnitude diagram of the SSA22 fields. $BV$ images are constructed by (2$B$+$V$)/3. The blue lines show the criteria described in the text. The red curves show the 4$\sigma$ error of the shallowest SSA22-Sb4 field. While all the objects satisfy the criteria is shown by the thick red dots, one-tenth of the other objects are shown by the thin black dots to avoid the confused appearance. The upper bound of the color corresponds to the 1$\sigma$ detection limit of the $BV$ image.}
\label{fig6}
\end{figure*}

\figurenum{7}
\begin{figure*}
\includegraphics[width=16cm]{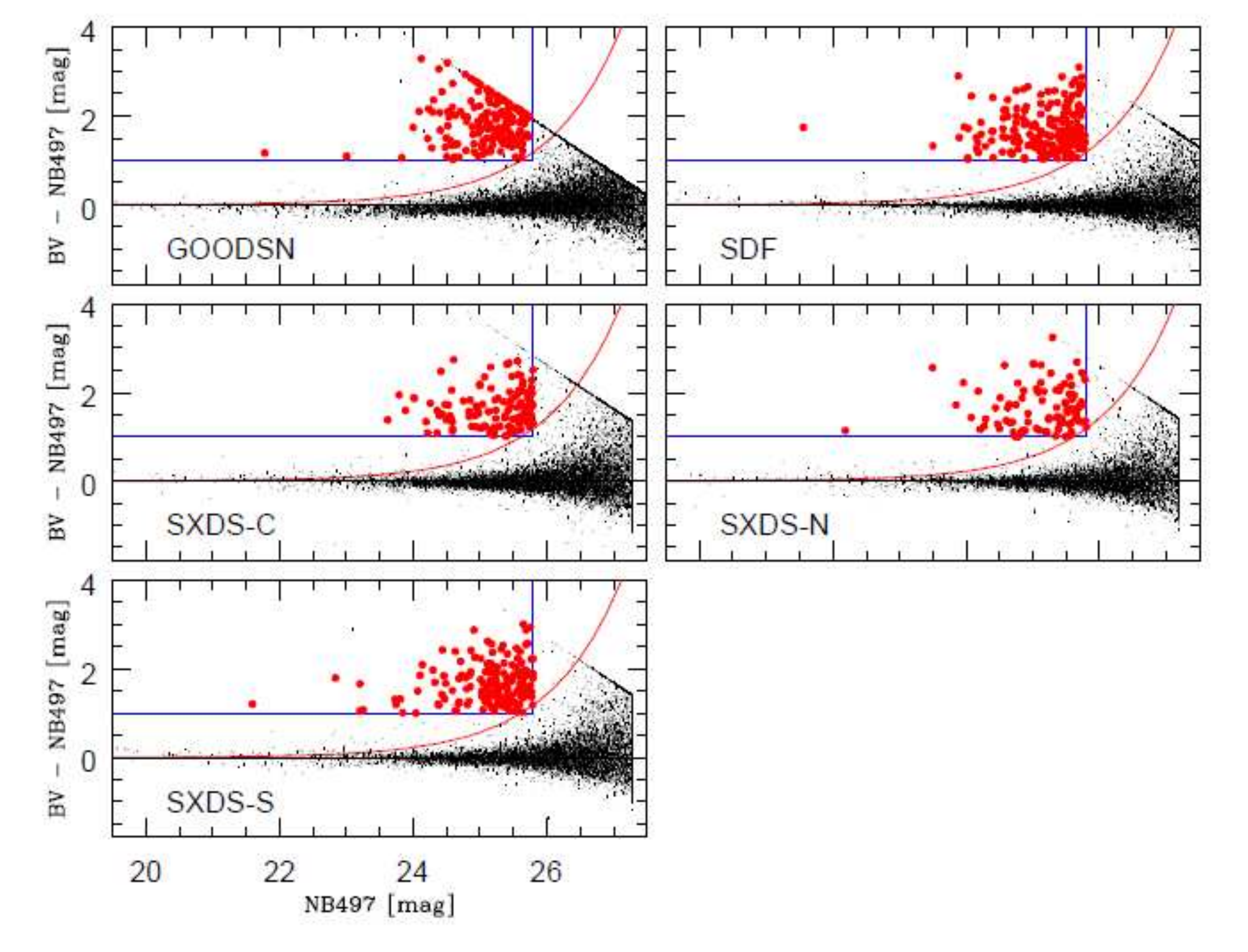}
\caption{Same as Fig.6 but for the general fields.}
\label{fig7}
\end{figure*}

 In total, we identified the 1394 and 767 emission line objects in the SSA22 and blank fields, respectively.  The $NB497$ limiting magnitude corresponds to $\approx 1.8$$\times 10^{-17}$ erg s$^{-1}$ cm$^{-2}$ or the luminosity of $\approx 1.5$$\times 10^{42}$ erg s$^{-1}$ at $z$=3.09. The numbers and the area for each field (without overlapping) are summarized in Table 2. Based on the previous studies for SSA22-Sb1, including the spectroscopic observations, we emphasize that the sample is quite robust as the $z$=3.1 Ly$\alpha$ emitters and not contaminated by the foreground objects more than $\sim 1\%$ in the fraction. This is also confirmed by the recent spectroscopic data for 100 sources in Sb1 (Yamada et al., in preparation) as well as the on-going observations for Sb2-Sb7 for a few hundred sources. We show the number counts of the $NB497$, $B$, and $V$ band of the Ly$\alpha$ emitters in Fig.8-10.

%
%

\figurenum{8}
\begin{figure}
\includegraphics[width=75mm]{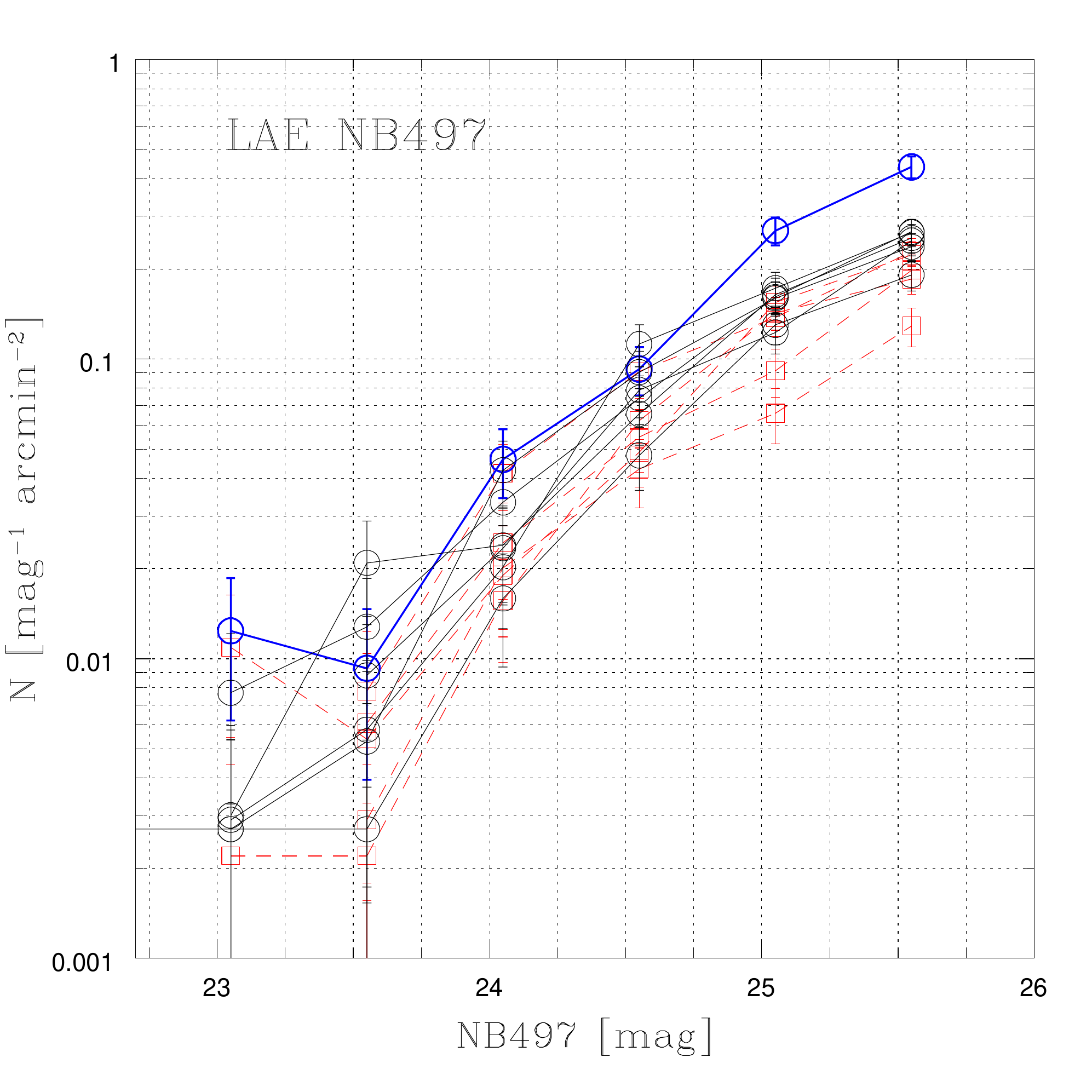}
\caption{The number counts of the selected Ly$\alpha$ emitters in each field. The blue line with the highest density is that for SSA22-Sb1. The five red dashed lines are for the general fields and the black lines are for the other SSA22 fields.}
\label{fig8}
\end{figure}

\figurenum{9}
\begin{figure}
\includegraphics[width=78mm]{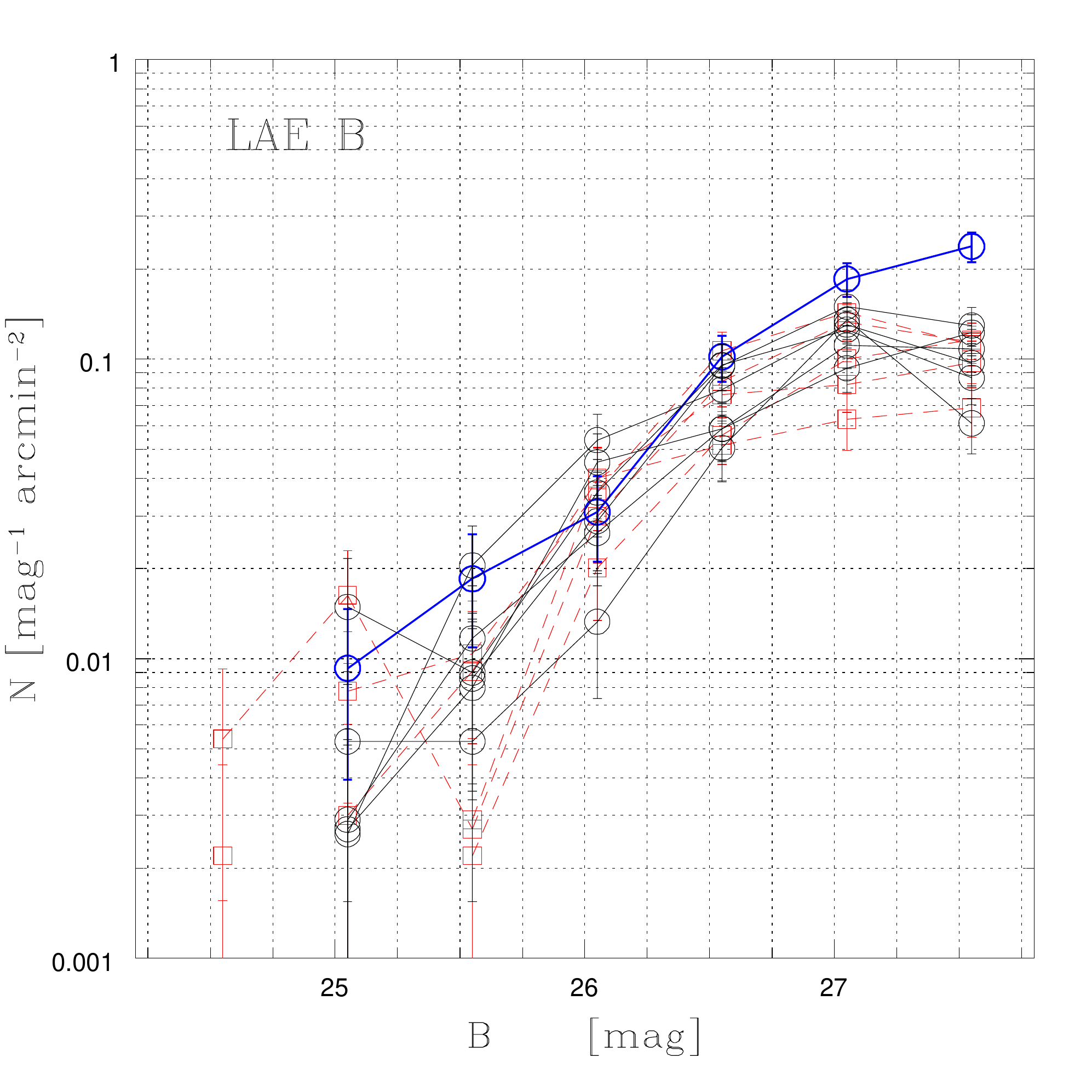}
\caption{Same as Fig.8 but in $B$ band.}
\label{fig9}
\end{figure}

\figurenum{10}
\begin{figure}
\includegraphics[width=78mm]{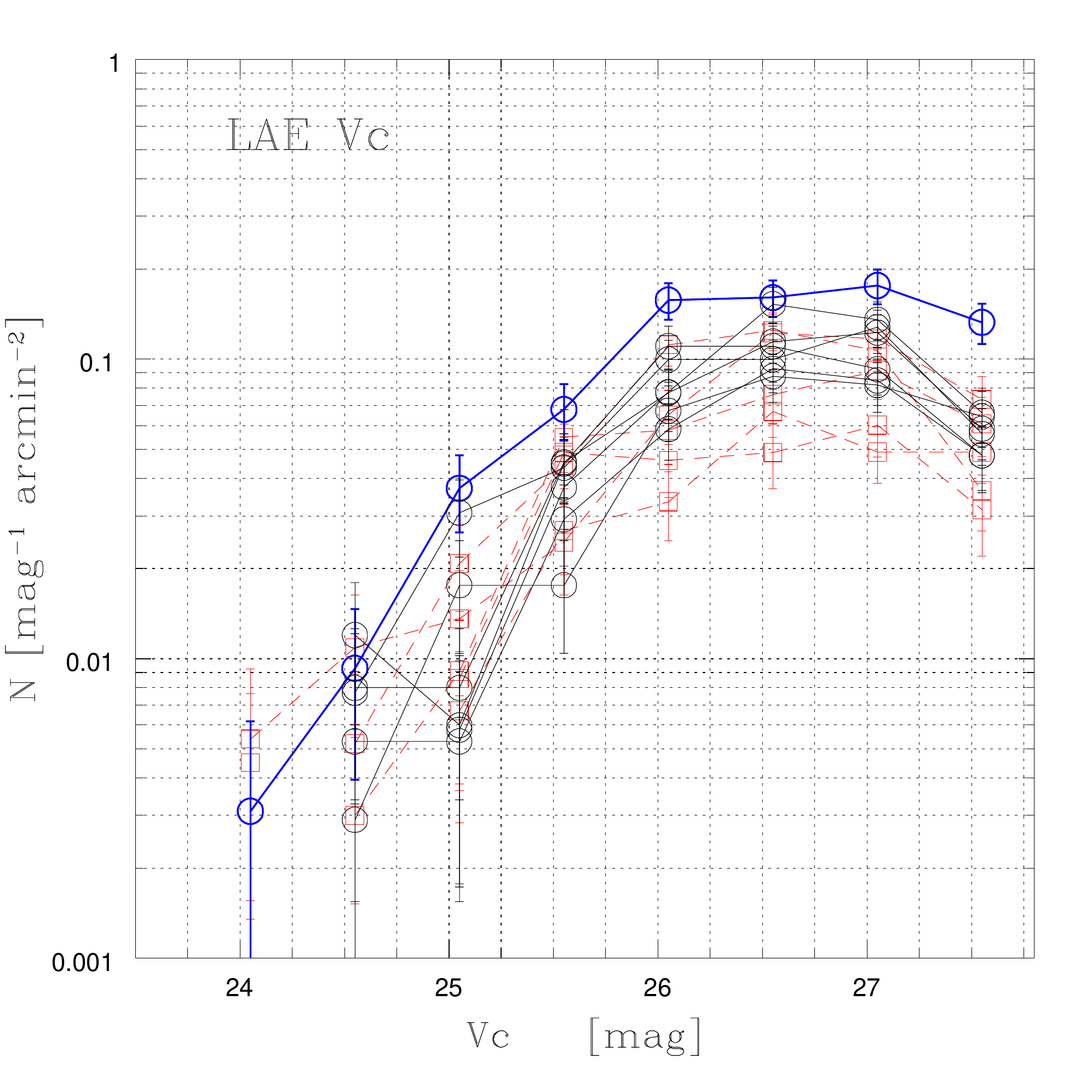}
\caption{Same as Fig.8 but in $V_c$ band.}
\label{fig10}
\end{figure}

\section{Results}

 The sky distributions of the $z$=3.1 Ly$\alpha$ emitters are shown in Fig.11-14. The separation of 1$^\circ$ corresponds to 109 Mpc in comoving scale at $z$=3.1. As shown in Table 2, the surface densities averaged in each Suprime Cam field of views of the five blank fields range from 0.13 (SXDS-N) to 0.26 (SDF) arcmin$^{-2}$ and the whole average of the five fields is 0.204$\pm 0.007$ armin$^{-2}$, which corresponds to be $1.1 \times 10^{-3}$ Mpc$^{-3}$. 

 We also calculated the average densities and the standard deviations with the randomly placed circular areas with various different diameters between 6$'$ and 20$'$. 
 The average values are nearly constant over the diameters and distribute within the ranges of 0.23-0.26 arcmin$^{-2}$ in SDF and 0.16-0.20 arcmin$^{-2}$ for the whole SXDS and GOODS-N while the standard deviation decreases as the diameter increases; it is $\approx 0.1$ arcmin$^{-2}$ at the diameter of $6'$ and $\approx 0.03$ arcmin$^{-2}$ at 20$'$. In this paper, we adopted 0.20 armin$^{-2}$ for the nominal average number density of the Ly$\alpha$ emitters selected by our criteria described above. 
 We also calculated the local smoothed number density with the Gaussian kernel with $\sigma$=1.$'$5 and the contours in Fig.11-Fig.14 show the smoothed densities of 1.5, 2.5, 3.5, 4.5, and 5.5 times the nominal average density. The large-scale inhomogeneity in the SSA22 is clearly seen in Fig.11 and the number density excess in the SSA22-Sb1 is evident. The highest peak with $\approx 6$ times the average is seen in the SSA22-Sb1 field at ($\alpha$, $\delta$) $\approx$ (334.$^\circ$35, 0.$^\circ$3), right at the position of the local "peak" discussed in Steidel et al. (2000) (north-east one in their Fig.10). There is only one small peak with $\sim 3.5$ times the average in SDF but none in other general fields. The highest peak in the SSA22-Sb1 is thus conspicuous very much. 

 The densities of the SSA22 Suprime Cam fields range from 0.20 to 0.44 arcmin$^{-2}$ (Table 2) and the whole average is 0.28 arcmin$^{-2}$. The high surface density of the SSA22-Sb1 field is prominent and more than twice higher than the average of the general blank fields. There is also a large under dense region over $\approx 30$ Mpc (in comoving scale) in the eastern half of SSA22-Sb6 at around ($\alpha$, $\delta$)$\approx$(333.$^\circ$7, 0.$^\circ$6) while the surface density is  high in the western half of SSA22-Sb5 at around $\approx$(333.$^\circ$7, 0.$^\circ$2). The whole average of the SSA22 fields omitting SSA22-Sb1 is 0.25 armin$^{-2}$, which is still higher than the average but a similar value to the SDF which is the densest among them. 

%
%
\figurenum{11}
\begin{figure*}[!p]
\includegraphics{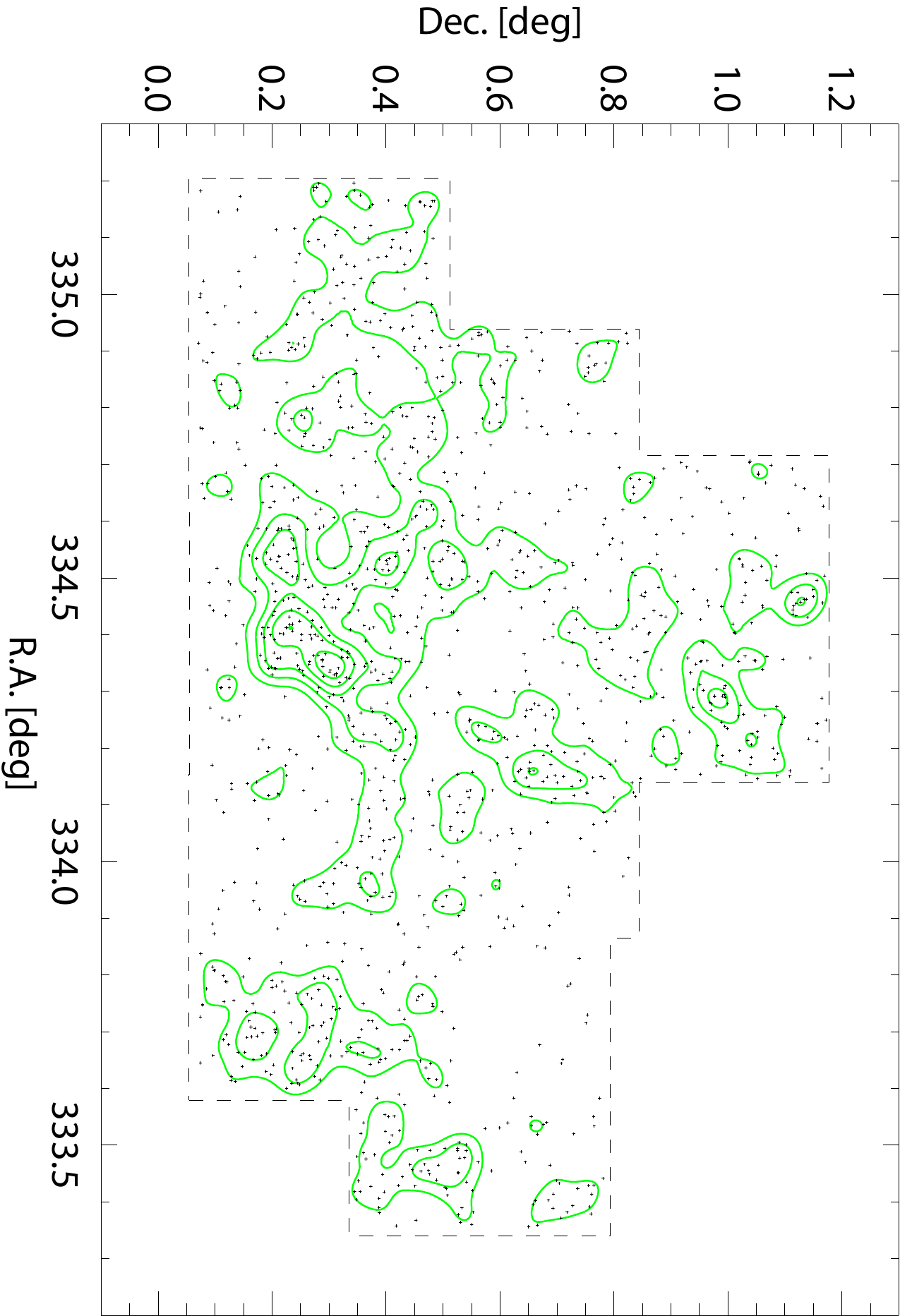}
\caption{The sky distributions of the Ly $\alpha$ emitters detected in the SSA22 fields.}
\label{fig11}
\end{figure*}

\figurenum{12}
\begin{figure}[!h]
\includegraphics[width=8cm]{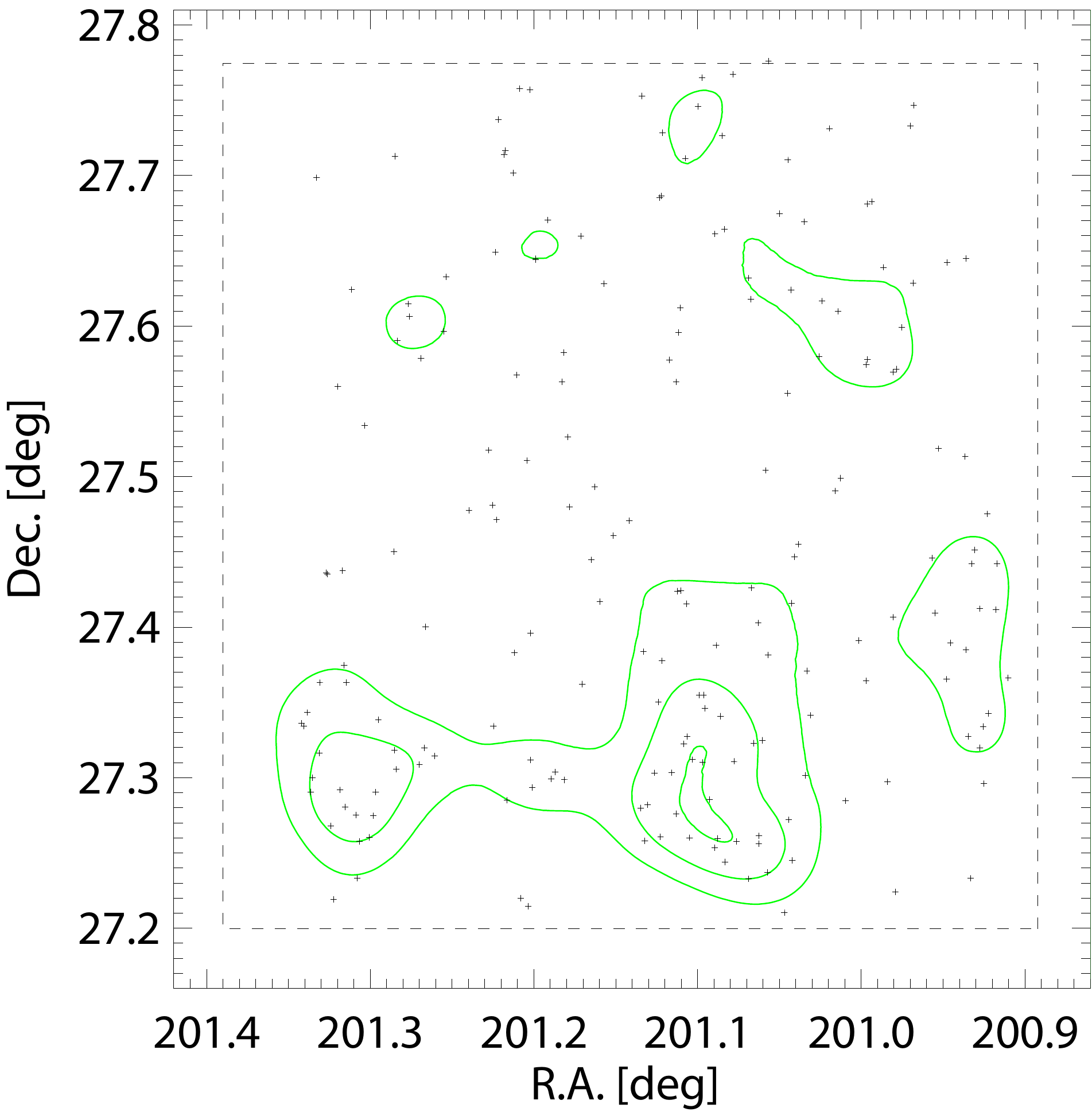}
\caption{The sky distributions of the Ly $\alpha$ emitters detected in the SDF field.}

\label{fig12}
\end{figure}

\figurenum{13}
\begin{figure}[!h]
\includegraphics[width=8cm]{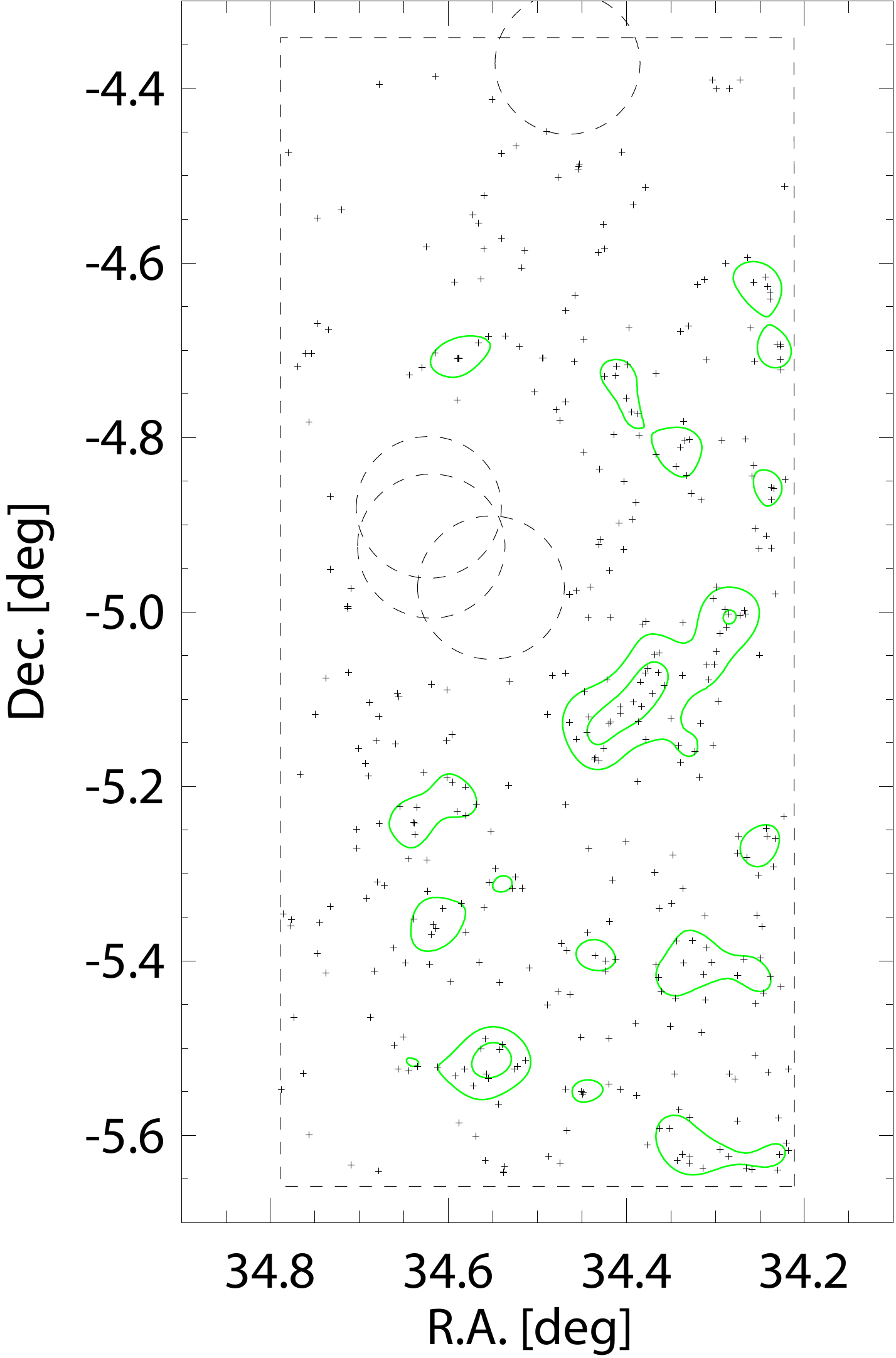}
\caption{The sky distributions of the Ly $\alpha$ emitters detected in the SXDS fields. The large ciecles of dashed lines indicate the area of the very bright stars which are omitted from the analysis.}
\label{fig13}
\end{figure}

\figurenum{14}
\begin{figure}[!h]
\includegraphics[width=8cm]{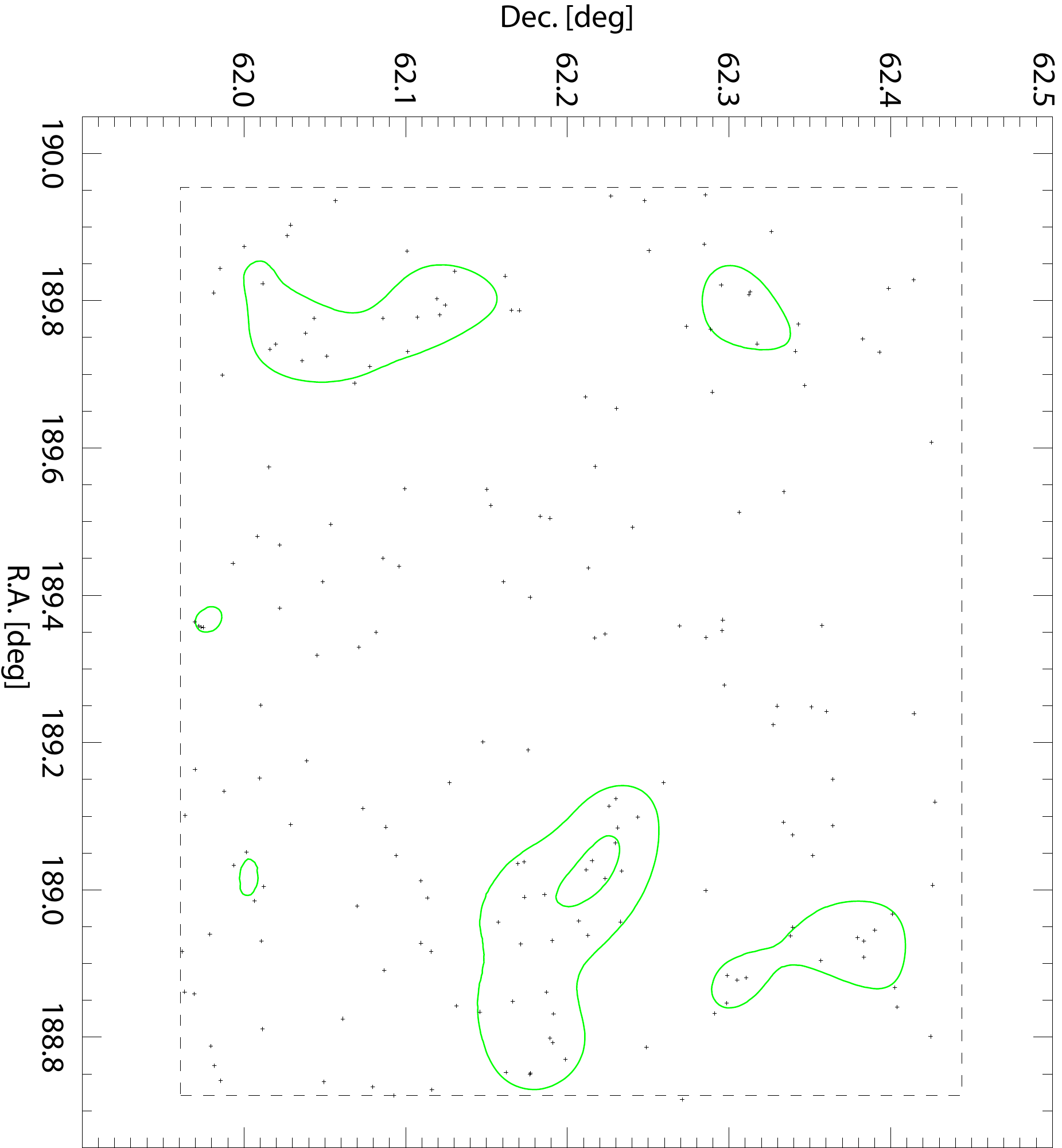}
\caption{The sky distributions of the Ly $\alpha$ emitters detected in the GOODS-N field.}
\label{fig14}
\end{figure}

 We also compare our results with the previous survey of Ly$\alpha$ emitters at $z \sim 3$. Ouchi et al. (2008) observed the $z$=3.1 Ly$\alpha$ emitters in SXDS by the slightly different narrow-band filter $NB503$ and detected 356 objects in 0.98 deg$^2$ brighter than $NB503=25.3$ and $V-NB503 > 1.2$, which yields the surface number density of 0.099$\pm 0.005$ arcmin$^{-2}$. While the selection criteria is slightly different, we also counted the number of our Ly$\alpha$ emitters with $NB497 < 25.3$ and $BV-NB497 > 1.2$. We found 309 and 530 Ly$\alpha$ emitters in the five general fields (1.04 deg$^2$) and SSA22 (1.38 deg$^2$), respectively. The surface density of our sample in the general fields is thus 0.083$\pm 0.004$ arcmin$^{-2}$ which is only slightly smaller than the value in Ouchi et al. (2008). On the other hand, the average surface density in the whole SSA22 field and that in SSA22-Sb1 are 0.11 arcmin$^{-2}$ and 0.18 arcmin$^{-2}$, respectively, which is similarly high even if we compare with the average density of the$NB503$ emitters in Ouchi et al. (2008) to the magnitude and equivalent width limit. 

 Gronwall et al. (2007) also searched the Ly$\alpha$ emitters at $z$=3.1 down to the magnitude limit of $\approx 25.4$ and the equivalent width limit of $\sim 20$\AA\ in 0.28 deg$^2$ and found 167 emitters, or 0.16 arcmin$^{-2}$. As the width of the narrow-band filter is somewhat more different (50 \AA\ in Gronwall et al.), the direct comparison is rather difficult. If we apply the similar selection threshold and corrected the 1.5 times volume difference in a unit area, we found 0.16 arcmin$^{-2}$ for the general fields, and 0.24 and 0.34 arcmin$^{-2}$ in the whole SSA22 and SSA22-Sb1, respectively. Thus our result is reasonably consistent with the previous results in terms of the number density in the general fields and the very large density excess in the SSA22 fields is robust. 

 The color-magnitude diagrams of Fig.6 and Fig.7 show that the observed narrow-band color excess of the emitters ranges from $BV-NB497$=1 (selection limit) to at least 3.5, which corresponds to the equivalent width of the Ly$\alpha$ emission between $\approx 190$\AA\ to 1900\AA\ in the observer's frame, or $\approx 50$ \AA\ to 500 \AA\ in the rest frame at $z$=3.09. Note that the emission line locates at the shoulder of the $V$-band. If we correct the line flux in the $V$ band, the $BV-NB497$ becomes even larger, especially for the objects with large equivalent width. The attenuation of the Ly$\alpha$ emissions by the inter-galactic medium (IGM) in the line of site ($\approx 20$\% at $z$=3) should also be corrected. 

 Interestingly, there are a large fraction of the objects whose uncorrected rest-frame Ly$\alpha$ equivalent width larger than $\sim 150$\AA\ , which corresponds to $BV-NB497$=2.3. For the constant continuous star-formation model with the 0.05 solar metallicity and Salpeter initial mass function with the upper and lower mass cut of 120 and 0.1 M$_\odot$, respectively, the expected equivalent width at the age larger than 10 Myr at the equilibrium phase is $\sim 150$ \AA\ without any reprocess of the Ly$\alpha$ and UV flux (Malhotra \& Rhoads 2002; Schaerer 2003). If the Ly$\alpha$ emission suffers from larger dust extinction due to the large optical path by scattering before escape, this becomes smaller. In the clumpy medium case (Finkelstein et al. 2007), on the other hand, the Ly$\alpha$ equivalent width may be enhanced at most $\sim 50$$\%$, it may be as large as $\sim 230$\AA\, which corresponds to $BV-NB497$=2.7. The {\it observed} $BV-NB497$ distribution of our sample spread over the value. This result suggests that the sample of Ly$\alpha$ emitter is powered by the photoionization with hotter stars, namely, younger or metal poor stars, or by other processes such as the collisional gas heating during the gravitational collapse or galactic superwind. 

 The distribution of the rest-frame equivalent width calculated from the color measured in 2$''$-diameter aperture is shown in Fig.15. We here corrected for both the $V$-band line contamination and the attenuation by IGM. We adopted the same value for the IGM attenuation factor, 0.81, as quoted in Ouchi et al. (2008) at $z$=3.1. The counts in the SSA22 fields from the objects whose $BV$ band flux is below the detection limit (i.e., only the lower limit of the equivalent width available) is indicated by the light magenta histogram. If we consider the whole range of the equivalent width, the distribution is not fitted very well by a single exponential law.  The best fitted decaying scale is 112.4$\pm$7.2 \AA\ for the whole SSA22 fields and 97.9$\pm$8.7 \AA\ for the whole sample of the general fields except for GOODS-N. The fraction of the objects with large equivalent width is larger in SSA22. On the other hand, if we fit them below 200\AA\ , the e-folding value is 73.68$\pm$30.59 \AA\ and 63.86$\pm$27.75 \AA\ , for the SSA22 fields and the general fields, respectively. The larger error is due to the small number of the bins. Fig.16 shows the same distribution of the equivalent width but in the cumulative manner. It is more clearly shown that the distribution in the SSA22 fields is more weighted to the larger equivalent-width objects, which is not due to the difference of the detection limit of the continuum bands.

 We may also compare the distribution of the equivalent width although the different continuum depth makes the direct comparison more difficult. Gronwall et al. (2007) found the e-folding scale 76$+11-8$\AA . This is smaller than our results for the full sample for the general fields, 97.9$\pm$8.7 \AA\ although they are not inconsistent very much considering the errors.

 Since the Ly$\alpha$ emission is typically more extended than the continuum (Hayashino et al. 2004), it may not be appropriate to calculate the equivalent width in the fixed circular aperture  when we are interested in the global process in the galaxies for the origins of the Ly$\alpha$ emission. While the fixed aperture should be used to evaluate local process like the photoionization in the star-forming region, the whole extent of the Ly$\alpha$ emission should be considered if we include the global scattering effects or the other process like shock heating by infalling or outflowing gas. We found that the equivalent width measured in the pseudo total aperture (i.e., SExtractor FLUX AUTO using the Kron aperture) is in average $\approx 1.6$ times larger than those measured with the 2$"$ diameter apertures. The equivalent-width distribution, which is already spread to the large value above 150\AA , is further stretched to show the large fraction of the large equivalent-width objects. This trend is seen more conspicuously for the emitters in the SSA22 fields. 

\section{Discussion}

 We have constructed a robust and homogeneous sample of the Ly$\alpha$ emitters at $z$=3.1 around the SSA22 high-density peak as well as the blank fields. The primary result is that the high-density region originally discovered by Steidel et al. (1998) is now characterized as the very significant structure. It is the most conspicuous density peak in the survey area in total of 2.42 deg$^2$, or in the comoving volume of $1.8 \times 10^6$ Mpc$^3$. No other region shows the density larger than 3.5 times the average of the blank fields. For the number density, the SSA22 high-density region is likely to be rarer than the present-day clusters of galaxies with $L_X$(0.5-2 keV) $\sim 10^{44}$ erg s$^{-1}$ (B{\"o}hringer et al. 2002), which is about the half luminous of the Coma cluster. 

 From Fig.11, the extension of the density peak is $\sim 5$ arcmin $\times 10$ arcmin, or $\sim 3 \times 6$ Mpc at $z$=3.1 in the proper length, if we take the contour of the 3.5 times the average as the boundary. The size is comparable with the extension of the X-ray gas of the rich cluster (e.g., $\sim 3$Mpc for Coma cluster). The overdensity, $\delta N$/$N_0$, of the Ly$\alpha$ emitter in the region is also found to be very large. The density inside the $\sim 5 \times 10$ arcmin$^2$ area is $\sim 1.0$ arcmin$^{-2}$, which is 5 times the average density in the general fields, 0.20 arcmin$^{-2}$. This overdensity is $\approx 10 \times$ of the standard deviation of the Ly$\alpha$ emitters in the general field, 0.09 arcmin$^{-2}$. The structure should correspond to the extremely rare peak which must be a progenitor of the very massive structure if the clustering of the Ly$\alpha$ emitters obey the clustering of the dark matter halos. From all these above, we can robustly call the structure a `protocluster'.

 We now evaluate the significance of the overdensity of the high density region of the Ly$\alpha$ emitters at some different scales by comparing with the prediction of the structure formation models.
 The area of 700 arcmin$^2$, a Suprime Cam field of view, corresponds to the comoving volume of 1.4$\times 10^5$ Mpc$^3$, with the depth sampled by the narrow band, 59Mpc. The distribution of the mass or the number-density fluctuation of the galaxies is well approximated by the Gaussian distribution at $z$=3.1 at this scale. The underlying mass fluctuation can be roughly estimated assuming the standard $\Lambda$CDM model with the linear approximation. Adopting the normalization of the power spectrum $\sigma_8$=0.81 (Komatsu et al. 2011) and the approximation for the mass variance with the top-hat window, $\sigma (R) = \sigma_8 (R/r_8)^{-\beta}$ (Mo et al., 2010) with $\beta \approx 0.7$ (CDM shape parameter $\Gamma=0.2$), and the linear growth rate in the non-zero cosmological constant (Carroll et al. 1992), $\approx 3.3$ at $z$=3.1, we obtain the $\sigma(R)$=$\langle$($\delta\rho$ / $\rho_0$)$^2$$\rangle$$^{1/2}$ at the corresponding comoving scale ($R \approx 32$ Mpc) at $z$=3.1 is $\approx$ 0.11. 
 The observed density variation among the five general fields, ($\delta N$ / $N_0$)$_{\rm LAE}$=$-0.38$ (SXDS-N) to $+$0.21 (SDF) is comparable with the value, if we consider that the Ly$\alpha$ emitters traces mass distribution with the linear biasing factor $b \sim 2-6$, as obtained in the previous study of Ly$\alpha$ emitters at $z=2-5$ from the analysis of the two-point correlation functions (Shimasaku et al. 2003; Gawiser et al. 2007 ; Guaita et al. 2010). On the other hand, the density excess of the SSA22-Sb1 with ($\delta N$ / $N_0$)$_{\rm LAE}$$=1.13\pm0.01$ even at this scale is unexpectedly large, more than $\approx 10$ times of the mass fluctuation, which implies that the field is the very rare, $>5$$\sigma_{\rm LAE}$ peak with $b$=2 (Gawiser et al. 2007). 

%
%
\figurenum{15}
\begin{figure}[!h]
\includegraphics[width=78mm]{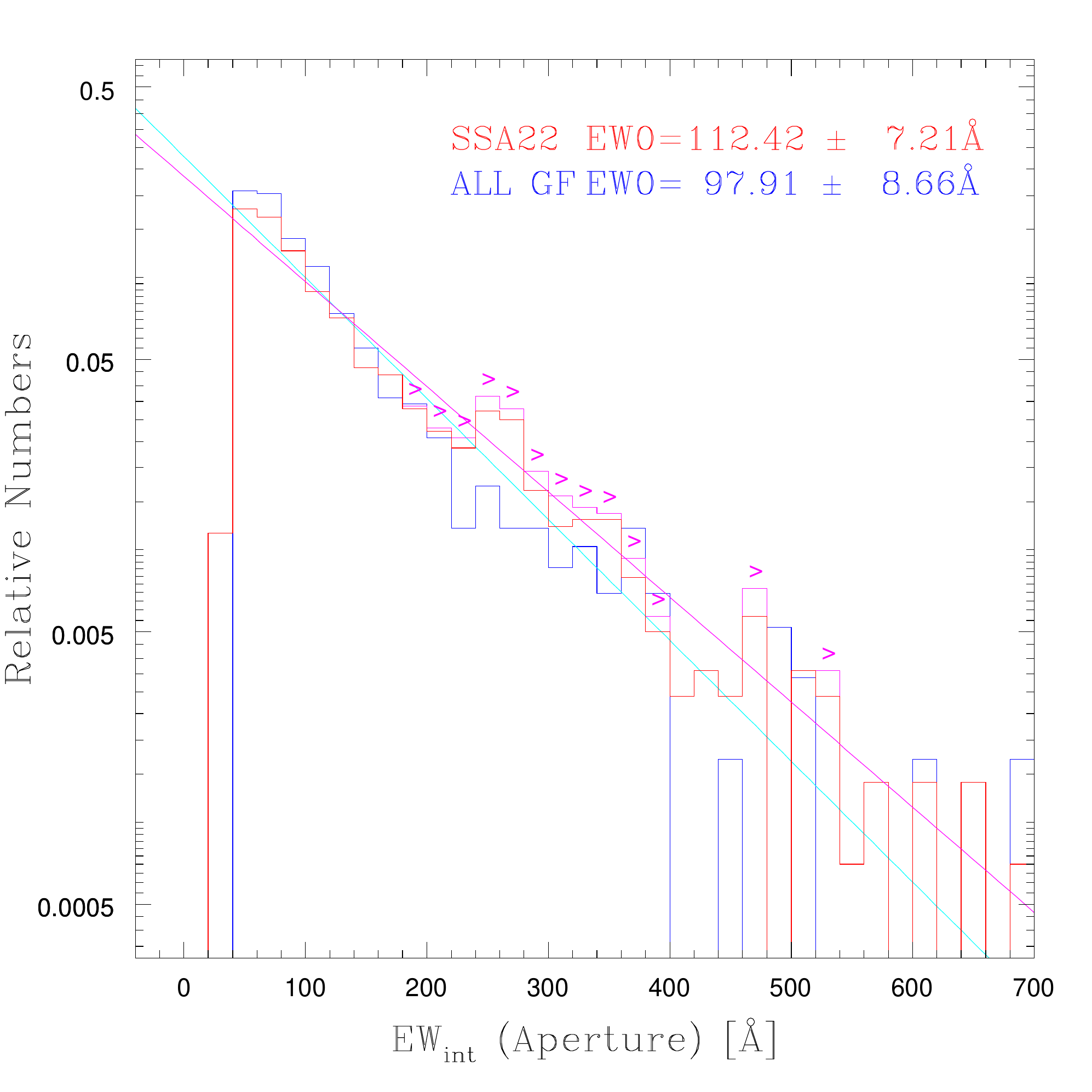}
\caption{The distributions of the equivalent width calculated from the $2"$-diameter aperture magnitude values. The red histogram is that of the entire SSA22 fields and the magenta lines with '$>$' show the fraction of the objects whose $BV$-band continuuum fainter than the upper limit and the obtained equivalent width values are only lower limit. The blue histogram is that of the SDF and SXDS fields. The lines are the fitted exponential distribution.}
\label{fig15}
\end{figure}

\figurenum{16}
\begin{figure}[!h]
\includegraphics[width=78mm]{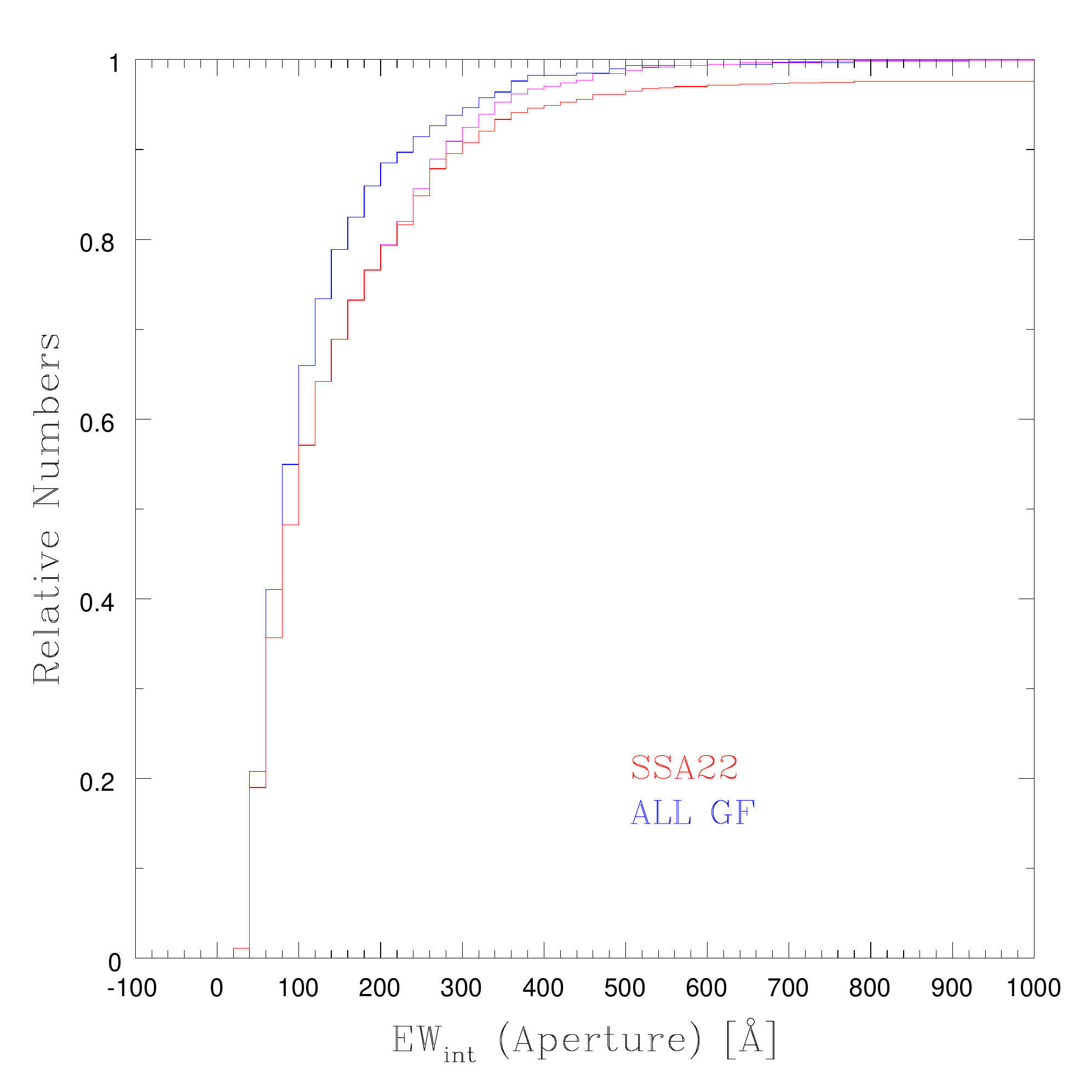}
\caption{Same as in Fig.15 but presented in the cumulative manner. The magenta line shows the fraction of the lower-limit values of the equivalent width.}
\label{fig16}
\end{figure}

 In Table 3, we also compared the observed overdensity with the expected mass fluctuation not only for the smaller area with 100 arcmin$^2$ that contains the local density peak of the emitters, but also for the entire seven fields of SSA22. Note that the average density of the general blank fields is not changed very much with the observed area (Sec.3) and it is reasonable to assume the constant average surface density over these scales. The 100 arcmin$^2$ area centered at ($\alpha$, $\delta$)=(334.392, 0.250), which corresponds to the SSA22a in Steidel et al. (2000) contains 89 emitters, which corresponds to the overdensity of 3.36$\pm 0.03$ while the 1$\sigma$ mass fluctuation is 0.18. We assumed the mass fluctuation at this scale is still represented by the $\sigma (R)$ evaluated as above, although the entire distribution may be deviated from the Gaussian at this scale at $z$=3.1. The peak is confirmed to be very significant, 19 times of the average mass fluctuation.  At the larger scale of the entire SSA22 Sb fields, the overdensity is observed to be 0.37$\pm0.01$ while the expected mass fluctuation is 0.07 at this scale. With $b$=2, it is the 2.6$\sigma$ peak at the dimension of $\approx 100$Mpc scale. The SSA22 field is a very significant high density region in the universe at the scale of the Great Wall (Geller and Huchra 1989). 

  Furthermore, we compared the overdensity with the distribution of the model galaxies in the cosmological simulation, too. We use the public semi-analytic model galaxy catalogs at $z$=3.06 of the Millennium simulation (De Lucia et al. 2006). The catalog gives the distribution of galaxies within the 500$h^{-1}$ Mpc cube and the model parameters for the star-formation rate (SFR) and the mass weighted age are available. It is not easy to make the sample of the model galaxies which exactly represents the observed Ly$\alpha$ emitters since the Ly$\alpha$ emission may be diverse not only by the condition of ionization and excitation of the gas but also by dust extinction and resonance scattering. We here chose the sample of star-forming galaxies which may crudely mimic the observed galaxies and investigate their general properties. We first make the samples of the model galaxies simply divided by their SFR, those with SFR $>$ 18 M$_\odot$ yr$^{-1}$ (Sample A),  10 M$_\odot$ yr$^{-1}$ $<$ SFR $<$ 18 M$_\odot$ yr$^{-1}$ (Sample B), and 6.5 M$_\odot$ yr$^{-1}$ $<$ SFR $<$ 10 M$_\odot$ yr$^{-1}$ (Sample C) in order to see how the fluctuation behaves as the SFR of the model galaxies changes.  The ranges are chosen so that the number density is the same as the observed value in the general blank fields. The observed rest-frame ultra-violet continuum flux of the Ly$\alpha$ emitters, $V$=25-27 mag, corresponds to the SFR of a few  to $\approx10$ M$_\odot$ yr$^{-1}$ at $z$=3.1 without reddening correction (Matsuda et al. 2004). Considering that the Ly$\alpha$ luminosity may vary both with the SFR and age, we also make a sample (Sample D) of the model galaxies which satisfy the condition  $log$ $\tau$ $<$ A / SFR where $\tau$ is the age and $A$ is the constant to match their number density to the observed value. Model galaxies either large SFR or young age are selected by the criteria.

 We then randomly put the boxes which correspond to the 700 arcmin$^2$ rectangular Suprime Cam field of view (1000 boxes), or the 100 arcmin$^2$ square region (5000 boxes) both with the depth of 59 Mpc, in the volume of the Millennium simulation to obtain the distribution of the counts in the boxes. The each distribution is fitted by a single Gaussian function in order to evaluate the standard deviation, $\sigma_{\rm model}$ of the model galaxies in the sample. While the distribution for the 700 arcmin$^2$ sample can be fitted by the Gaussian function very well from low to high overdensity values, the small tails of the high overdensity are seen for the 100 arcmin$^2$ samples, especially for the Sample A with the large SFR, which can be better represented by the log-normal type probability distribution function. Table 4 summarizes the results. We found that the fluctuation becomes smaller for the sample of the lower SFR, which is reasonable if the objects with higher SFR galaxies are associated with more massive haloes on average. But this trend is small and the standard deviation is not much different among the samples. Compared with these values, the observed overdensity is significantly large, 3.6-4.5 $\sigma_{\rm model}$ at the 700 arcmin$^2$ scale, and 8.4-9.1 $\sigma_{\rm model}$ at the 700 arcmin$^2$ scale. 

 From those above, we conclude that the high density region of the SSA22 is the very rare density peak with 3-4$\sigma$ significance at the scale of $\approx 60$Mpc (diameter), and likely to be 2-3$\sigma$ at $\approx 100$Mpc. The overdensity is 8-10 times of the standard deviation at the $\approx 30$Mpc scale. Note that there is no region with the similar overdensity of the model galaxies in the (500$h^{-1}$)$^3$ Mpc$^3$ volume of the Millennium simulation in our realizations (1000 boxes and 5000 boxes for the larger and the smaller scales, respectively).  

 The SSA22 structure is thus characterized to be the very important large-scale highest density region to study the earliest galaxy formation and evolution in the universe. There are indeed evidences that the active phenomena related with galaxy formation is enhanced over the large scale structure. It is important to note that the number density of the giant Ly$\alpha$ blobs with the size larger than 100 kpc is significantly high in the SSA22 area compared with the other general fields; Matsuda et al. (2010) reported the detection 12 such giant blobs in the SSA22 fields (in 1.38 deg$^2$) while only 2 objects in the general fields (1.04 deg$^2$). Six objects are distributed in the SSA22-Sb1 field, and of them, five are clustered near the density peak of the distribution of Ly$\alpha$ emitters, although they seem to avoid the very central region (See Fig.3 of Matsuda et al. 2010). These giant Ly$\alpha$ blobs are likely to be the massive galaxies in their forming phase (Yamada 2009; Uchimoto et al. 2008; 2011). While the formation of the massive galaxies in the densest part might have occured ralatively earlier or faster, as the density peak of the stellar massive objects coinsides with that of the Ly$\alpha$ emitters (Uchimoto et al. 2011), it is more active in the surrounding volume at the epoch.  Tamura et al. (2009) also reported the significant clustering of the bright sub-mm galaxies in the SSA22 field and they show positive cross correaltion with the sky distribution of the Ly$\alpha$ emitters. Geach et al. (2005; 2007) indeed found that more than a few Ly$\alpha$ blobs in Sb1 field are the bright sub-mm or infrared sosurces. Lehmer et al. (2009) and Webb et al. (2009) also show that the fraction of AGN is also high in the SSA22 field and several Ly$\alpha$ blobs are asoociated with AGN. The SSA22 region can be {\it the Rosetta Stone} to understand the early galaxy formation as a function of the large-scale density of the universe.

\vspace*{5mm}

 We thank the staff of the Subaru Telescope for their assistance. This research is supported in part by the Grant-in-Aid 20450224 for Scientific Research of the Ministry of Education, Science, Culture, and Sports in Japan. Data analysis was in part carried out on common use data analysis computer system at the Astronomy Data Center, ADC, of the National Astronomical Observatory of Japan.


\clearpage

\tablenum{1}

\begin{table}
\begin{center}
\caption{Summary of the Observed Objects\label{tab1}}
\begin{tabular}{ccccccc}
\tableline\tableline
Field &   R.A.   & Dec.     &  Filter &  FWHM   &  5$\sigma$ limit  &  Exp. Time \\ 
      & J2000.0  & J2000.0  &         &  arcsec &  AB mag.          &  hours    \\
\tableline
SSA22-Sb1 & 22:17:34.0 & +00:17:01.0 &  $NB497$ & 1.0 & 26.33 & 7.0  \\
          &            &             &    $B$   & 1.0 & 26.49 & 1.5  \\
          &            &             &    $V$   & 1.0 & 26.69 & 1.5  \\
SSA22-Sb2 & 22:16:36.7 & +00:36:52.0 &  $NB497$ & 1.0 & 26.25 & 5.5  \\
          &            &             &    $B$   & 1.0 & 26.53 & 1.0  \\
          &            &             &    $V$   & 1.0 & 26.36 & 1.0  \\
SSA22-Sb3 & 22:18:36.3 & +00:36:52.0 &  $NB497$ & 1.0 & 26.30 & 5.5  \\
          &            &             &    $B$   & 1.0 & 26.81 & 1.0  \\
          &            &             &    $V$   & 1.0 & 26.55 & 1.0  \\
SSA22-Sb4 & 22:19:40.0 & +00:17:00.0 &  $NB497$ & 1.1 & 25.85 & 5.5  \\
          &            &             &    $B$   & 1.1 & 26.31 & 1.0  \\
          &            &             &    $V$   & 1.1 & 26.21 & 1.1  \\
SSA22-Sb5 & 22:15:28.0 & +00:17:00.0 &  $NB497$ & 1.0 & 26.11 & 5.5  \\
          &            &             &    $B$   & 1.0 & 26.45 & 1.0  \\
          &            &             &    $V$   & 1.0 & 26.37 & 1.0  \\
SSA22-Sb6 & 22:14:30.7 & +00:33:52.0 &  $NB497$ & 1.0 & 26.27 & 5.5  \\
          &            &             &    $B$   & 1.0 & 26.83 & 1.3  \\
          &            &             &    $V$   & 1.0 & 26.58 & 1.3  \\
SSA22-Sb7 & 22:17:42.7 & +00:56:52.0 &  $NB497$ & 1.0 & 26.15 & 5.5  \\
          &            &             &    $B$   & 1.0 & 26.62 & 1.0  \\
          &            &             &    $V$   & 1.0 & 26.44 & 1.0  \\
SDF       & 13:24:39.0 & +27:29:26.0 &  $NB497$ & 1.0 & 26.53 & 7.2  \\
          &            &             &    $B$ & 1.0 & 27.82 & 9.9  \\
          &            &             &    $V$ & 1.0 & 27.25 & 5.7  \\
SXDS-N    & 02:18:00.0 & -05:25:00.0 &  $NB497$ & 1.0 & 26.19 & 4.9  \\
          &            &             &    $B$ & 1.0 & 27.56 & 5.5  \\
          &            &             &    $V$ & 1.0 & 27.08 & 1.4  \\
SXDS-C    & 02:18:00.0 & -05:00:00.0 &  $NB497$ & 1.0 & 26.28 & 5.3  \\
          &            &             &    $B$ & 1.0 & 27.51 & 5.8  \\
          &            &             &    $V$ & 1.0 & 27.29 & 1.6  \\
SXDS-S    & 02:18:00.0 & -05:00:00.0 &  $NB497$ & 1.0 & 26.28 & 4.8  \\
          &            &             &    $B^b$ & 1.0 & 27.60 & 5.5  \\
          &            &             &    $V^b$ & 1.0 & 27.24 & 1.8  \\
GOODS-N   & 12:37:23.6 & +62:11:31.0 &  $NB497$ & 1.1 & 26.55 & 10.0 \\
          &            &             &    $B$   & 1.1 & 26.68 & 2.3  \\
          &            &             &    $V$   & 1.1 & 26.15 & 1.0  \\
\tableline
\end{tabular}
\end{center}
\end{table}


\clearpage

\tablenum{2}

\begin{table}
\begin{center}
\caption{Numbers of the Detected Ly$\alpha$ Emitters\label{tab2}}

\begin{tabular}{cccc}
\tableline\tableline
Field &  Area       &   N$_{\rm LAE}$   &  Surface Density \\
      &  arcmin$^2$ &                   &  arcmin$^{-2}$ \\
\tableline
SSA22-Sb1 &   647.46       &    281          &    0.434          \\
SSA22-Sb2 &   683.70       &    166          &    0.243          \\
SSA22-Sb3 &   669.29       &    179          &    0.267          \\
SSA22-Sb4 &   750.49       &    202          &    0.269          \\
SSA22-Sb5 &   694.61       &    201          &    0.289          \\
SSA22-Sb6 &   753.14       &    148          &    0.197          \\
SSA22-Sb7 &   781.16       &    217          &    0.278          \\
SDF       &   769.44       &    196          &    0.255          \\
SXDS-C    &   656.31       &    123          &    0.187          \\
SXDS-N    &   697.39       &     92          &    0.132          \\
SXDS-S    &   735.75       &    171          &    0.232          \\
GOODS-N   &   897.16       &    185          &    0.206          \\
\tableline
\end{tabular}
\end{center}
\end{table}

\tablenum{3}

\begin{table}
\begin{center}
\caption{Significance of the overdensity compared with mass fructuation \label{tab3}}

\begin{tabular}{cccccc}
\tableline\tableline
Field &  Area        & N$_{\rm LAE}$   &  $\delta_{\rm LAE}$ & $R_{\rm th}$\footnote{The equivalent radius of the top-hat sphere for the volume.} &  $\sigma_{\rm mass}$ \\
      &  arcmin$^2$  &                 &                     &   Mpc      &                      \\
\tableline
SSA22 Sb1-Sb7 &   4980   &   1394          &    0.37$\pm$0.01  & 62     & 0.07 \\
SSA22 Sb1     &    647   &    281          &    1.13$\pm$0.01  & 31     & 0.11 \\
Sb1   peak    &    100   &     89          &    3.36$\pm$0.03  & 17     & 0.18 \\
\tableline
\end{tabular}
\end{center}
\end{table}


\tablenum{4}

\begin{table}
\begin{center}
\caption{Significance of the overdensity compared with the semi-analytic model\label{tab4}}

\begin{tabular}{cccccccc}
\tableline\tableline
Sample &  SFR range     & N$_{\rm 700}$$^a$  &  $\sigma_{\rm 700}$\footnote{The average numbers and the Gaussian standard deviations for the distribution of the counts in the box that corresponds to the  700 arcmin$^2$ area.} & Peak height (Sb1) &N$_{\rm 100}$$^b$ &  $\sigma_{\rm 100}$\footnote{The average numbers and the Gaussian standard deviations for the distribution of the counts in the box that corresponds to the  100 arcmin$^2$ area.} & Peak height (peak) \\

      &  M$_\odot$ yr$^{-1}$  &         &         &        &            \\
\tableline
Sample A &   $>18$   &   144   &    0.31 & 3.6 & 19 & 0.40 & 8.4  \\
Sanple B &   10-18   &   133   &    0.27 & 4.2 & 18 & 0.39 & 8.6  \\
Sample C &   6.5-10  &   151   &    0.25 & 4.5 & 21 & 0.37 & 9.1  \\
Sample D &  Age SFR  &   142   &    0.27 & 4.2 & 20 & 0.40 & 8.4 \\
\tableline
\end{tabular}
\end{center}
\end{table}


\end{document}